\UseRawInputEncoding
\documentclass[a4paper,fleqn,usenatbib]{mnras}
\usepackage{newtxtext,newtxmath}

\usepackage[T1]{fontenc}
\usepackage{ae,aecompl}

\usepackage{graphicx}	
\usepackage{amsmath}	
\numberwithin{equation}{section}

\title[Metals and a Search for Molecules in the Distant Universe]{Metals and a Search for Molecules in the Distant Universe: Magellan MIKE Observations of sub-DLAs at $2<z <3$}

\author[S. Poudel et al.]{
Suraj Poudel,$^{1,5}$\thanks{E-mail: spoudel@email.sc.edu, suraj.poudel@pucv.cl}
Varsha P. Kulkarni,$^{1}$
Debopam Som,$^{2}$
C\'eline P\'eroux$^{3,4}$
\\
$^{1}$University of South Carolina, Dept. of Physics and Astronomy, 712 Main Street, Columbia, SC 29208, USA\\
$^{2}$Space Telescope Science Institute, 3700 San Martin Drive, Baltimore, MD 21218, USA\\
$^{3}$European Southern Observatory, Karl-Schwarzschild-Strasse 2, 85748 Garching bei Munchen, Germany\\
$^{4}$Aix Marseille Univ, CNRS, CNES, LAM, Marseille, France\\
$^{5}$Instituto de Fisica, Pontificia Universidad Catolica de Valparaiso, Av. Universidad 330, Curauma, Valparaiso, Chile\\
}

\date{Accepted XXX. Received YYY; in original form ZZZ}

\pubyear{2020}

\begin{document}
\label{firstpage}
\pagerange{\pageref{firstpage}--\pageref{lastpage}}
\maketitle

\begin{abstract}
We present abundance measurements of the elements Zn, S, O, C, Si and Fe for four sub-DLAs at redshifts ranging from z$=$2.173 to z$=$2.635 using observations from the MIKE spectrograph on the Magellan telescope to constrain the chemical enrichment and star formation of gas-rich galaxies. Using weakly depleted elements O, S, and or Zn, we find the metallicities after the photoionization corrections to be [S/H]=$-0.50\pm$0.11, [O/H]$>$-0.84, [O/H]=$-1.27\pm$0.12, and [Zn/H]=$+0.40\pm$0.12 for the absorbers at z$=$2.173, 2.236, 2.539, and 2.635, respectively. Moreover, we are able to put constraints on the electron densities using the fine structure lines of C II$^{\star}$ and Si II$^{\star}$ for two of the sub-DLAs. We find that these values are much higher than the median values found in DLAs in the literature. Furthermore, we estimate the cooling rate l$_{c}$=1.20$\times10^{-26}$ erg s$^{-1}$ per H atom for an absorber at z$=$2.173, suggesting higher star formation rate density in this sub-DLA than the typical star formation rate density for DLAs at similar redshifts. We also study the metallicity versus velocity dispersion relation for our absorbers. Most of the absorbers follow the trend one can expect from the mass versus metallicity relation for sub-DLAs in the literature. Finally, we are able to put limits on the molecular column density from the non detections of various strong lines of CO molecules. We estimate 3$\sigma$ upper limits of log N(CO,J=0)$<$13.87, log N(CO,J=0)$<$13.17, and log N(CO,J=0)$<$13.08, respectively, from the non-detections of absorption from the J$=$0 level in the CO AX 0-0, 1-0,  and 2-0 bands near 1544, 1510, and 1478{\AA}.

\end{abstract}




\section{Introduction}
Since molecular clouds, formed out of the cold neutral clouds are the birthplaces of stars, neutral gas is key to the formation of stars and galaxies. Absorption lines in quasar spectra offer a promising tool to measure the chemical properties of neutral gas in distant galaxies. Based on the values of neutral hydrogen column densities, the quasar absorption line systems are classified into different categories. Damped Lyman-alpha Absorbers (DLAs) and sub-Damped Lyman-alpha Absorbers (sub-DLAs) are especially important as they provide most of the neutral gas required for star formation  \citep*[e.g.][]{Peroux 2003, Nagamine et al. 2004a, Nagamine et al. 2004b, Wolfe 2006}. DLAs and sub-DLAs have high neutral hydrogen column densities log $N_{\rm H\, I} \ge 20.3$ and 19.0 $\le$ log $N_{\rm H\, I} < 20.3$, respectively. Unlike the emission-line technique which samples only the bright or star forming galaxies (SFGs), absorption-line technique selects galaxies independent of their brightness. In addition, DLAs/sub-DLAs are less affected by photoionization compared to the Lyman Limit Systems (LLSs). DLAs/sub-DLAs are thus better suited for estimating the metallicity and help to trace the cycling of metals and neutral gas in the interstellar medium (ISM) and the circumgalactic medium (CGM).\\

One can constrain the evolution of the comoving density of metals in the Universe by measuring the metallicities of DLAs/sub-DLAs with a wide range of redshifts spanning a variety of cosmic epochs \citep[e.g.][]{Kulkarni et al. 2007}. Such studies of sub-DLA/DLA metallicity evolution have been carried out extensively in the past \citep*[e.g.][]{ Kulkarni 2002, Prochaska et al. 2003a, Kulkarni et al. 2005, Kulkarni et al. 2007, Rafelski et al. 2012, Jorgenson et al.  2013, Som et al. 2013, Som et al. 2015, Quiret et al. 2016, Poudel et al. 2018, Poudel et al. 2020, Poudel S. 2020}. However, a great deal remains unknown about the role they play in star formation and galaxy formation. 
In fact, many sub-DLAs are known to be more metal rich than typical DLAs, and may represent a different population of galaxies \citep*[e.g.][]{Khare et al. 2007, Kulkarni et al. 2010}. Therefore, DLAs and sub-DLAs may have different nucleosynthetic histories, and possibly exhibit different star formation rates. The higher average metallicity of sub-DLAs also suggests that they may have higher dust content and thus provide better conditions for the formation of molecules than DLAs.\\

\citet{Wolfe 2004} reported that the star formation rates per unit area for DLAs are in the range of 10$^{-3}$ to 10$^{-2}$ M$_{\odot}$ yr$^{-1}$ kpc$^{-2}$. Their findings were based on measurements of C II$^{\star}$ absorption in 45 DLAs at redshifts $1.6 < z < 4.5$. However, these values remain largely unexplored for sub-DLAs. This is because most past studies in this field have focused on DLAs rather than sub-DLAs. It is interesting to note that sub-DLAs are in fact more abundant  compared to DLAs: the differential $N_{\rm H I}$ distribution rises with decreasing $N_{\rm H I}$ \citep[e.g.][]{Zafar et al. 2013}, and the number density of sub-DLAs integrated over their $N_{\rm H I}$ range exceeds that of DLAs \citep[e.g.][]{Noterdaeme et al. 2012}. In spite of this, sub-DLAs have been ignored in many past studies of element abundances. This is because sub-DLAs often (though not always)  have lower metal column densities in comparison to DLAs, which makes it necessary to have high spectral resolution and high signal-to-noise ratio (SNR) data to accurately determine their element abundances.  High spectral resolution is also needed to resolve the C II$^{\star}$ $\lambda$ 1336 line without blending with the nearby C II $\lambda$ 1334 line, which is usually strongly saturated. Moreover, study of the closely spaced molecular lines associated with the vibrational and rotational bands of H$_{2}$ and CO require high spectral resolution. The Magellan Inamori Kyocera Echelle (MIKE) spectrograph on the Magellan (Clay) telescope provides high enough resolution (R$\sim$22,000-28,000) to resolve the C II$^{\star}$ $\lambda$ 1336 as well as the molecular lines of CO.\\

Here, we present a study of the metallicities, electron densities, and cooling rate for a sample of 4 sub-DLAs observed at high resolution with the MIKE spectrograph on the Magellan (Clay) telescope. Moreover, we discuss the effect of photo-ionization on the derived abundances, search for the molecular lines and put limits in the column densities of CO. In section~\ref{sec:obs}, we present details of our observations for the sub-DLAs and describe the data reduction process. In section~\ref{sec:method}, we describe the technical aspects of spectroscopic measurements of absorption lines. The results for each of the sub-DLAs in our sample are presented in section~\ref{sec:result}. In section~\ref{sec:discussion}, we discuss various aspects of our results and compare them with the literature. Finally, in section~\ref{sec:conclusion}, we summarize our conclusions.

\section{Observations and data reduction}
\label{sec:obs}
Our sample consists of four absorbers with neutral hydrogen column densities ranging from log N$_{\rm H\, I}$ = 19.00 to 20.05, at redshifts 2.173 to 2.635 along the sight lines to three quasars. These quasars were observed with the MIKE spectrograph on the Magellan Clay telescope at Las Campanas observatory in Chile as a part of NOAO program 2010A-0499 (PI Kulkarni). Observations were carried out with the 1 $\times$ 5 arcsec$^{2}$ slit  which resulted in a spectral resolution of $\sim$22,000 with the red arm and $\sim$28,000 with the blue arm. While the blue arm has a wavelength coverage of 3400 to 4900 {\AA}, the red arm has a wavelength coverage of 4900 to 9400 {\AA}. The sightlines were observed in multiple exposures, to facilitate the rejection of cosmic rays. The details of the observations are summarized in Table~\ref{tab:sum_obs}.\\

The data reduction was performed using the standard MIKE pipeline reduction package. This code was developed by S. Burles, J. X. Prochaska and R. Bernstein and is written in IDL. The reduction package performs bias subtraction using the overscan region before flat-fielding the 2-dimensional data. It then extracts individual spectral orders from the sky-subtracted flats. Wavelength calibration is done using exposures of a Th-Ar comparison lamp (which were obtained before and after each of the science exposures). This is followed by the correction to heliocentric velocities and the conversion from air wavelengths to vacuum wavelengths. Individual Echelle orders were then extracted. Orders from multiple exposures were combined to reduce the effects of cosmic rays. Finally, the quasar continuum was fitted with a spline or Legendre polynomial  (usually of order 4 or 5), and the spectra were normalized by this continuum fit.\\ 

\section{Voigt profile fitting and abundance measurements}
\label{sec:method}
We used the program VPFIT\footnote{https://www.ast.cam.ac.uk/rfc/vpfit.html} v. 12.2 for fitting Voigt profiles to estimate column densities of all detected atoms and ions. VPFIT enables fitting of multi-component Voigt profiles convolved with the instrumental profiles using multiple iterations. Moreover, VPFIT allows the Doppler $b$ parameters and redshifts of the corresponding components to be tied together. All of our spectra have high enough resolution (10 to 14 km s$^{-1}$) to resolve the blending of metal lines. Although saturation can cause errors to the fitting results of individual lines, we reduced such errors by using multiple lines \citep[e.g.][]{Penprase et al. 2010}. Most of the metal lines used for fitting are outside the Lyman-$\alpha$ forest and thus not affected by blending with hydrogen lines.\\
 
Lyman-$\alpha$ lines were fitted by fixing the redshifts as determined from the profile fits to the metal-lines. The higher-order Lyman series lines were not covered in most cases. Even when they were covered, they were blended with the Lyman-$\alpha$ forest features, and thus could not be used. To estimate the uncertainties in the H I column densities, we compared a series of fitted profiles with the observed data to make sure that the differences between the data and the fitted profiles do not exceed 2$\sigma$ noise level. Using the H I and metal column densities, we next determined the abundances of all the observed elements following the standard definition,
\begin{equation}
 $[Y/H] = log ($\rm N_{Y}/N_{\rm H\, I}$) - log (Y/H)$_\odot$$
 \end{equation}
 where $N_{Y}$ and $N_{\rm H\, I}$ are the column densities for the element Y and neutral hydrogen, respectively. The last term log $\rm (Y/H)_{\odot}$ is the abundance of the element Y in the Sun. Solar photospheric values were taken from \citet{Asplund et al. 2009}. Finally, the rest-frame wavelengths and oscillator strengths of all the relevant transitions were adopted from \citet{Morton 2004} and \citet{Cashman et al. 2017}.



\begin{table*}
	\centering
	\caption{Summary of targets and observations}
	\label{tab:sum_obs}
	\begin{tabular}{ccccccc} 
		\hline
		\hline
		\parbox[t]{1.6in}{Quasar Name\par RA(J2000), Dec(J2000)\strut}  & z$_{\rm em}$ & z$_{\rm abs}$ & \parbox[t]{0.5in}{log $\rm N_{\rm H\, I}$ \par (cm$^{-2}$)\strut}  & \parbox[t]{0.5in}{Exposure  \par Time (s)\strut} & \parbox[t]{0.7in}{Wavelength \par Coverage({\AA})\strut} & \parbox[t]{0.8in}{Spectral\par Resolution (R)\strut}\\  
		\hline
		\hline

\parbox[t]{1.6in}{J1106-1731\par RA: 11:06:07.47; Dec: -17:31:13.50 \strut} &  2.572 & \parbox[t]{0.3in}{2.173 \par 2.539 \strut} & \parbox[t]{0.5in}{20.05$\pm$0.10 \par 19.00$\pm$0.12 \strut} & $7200$ & 3,500-9,400 & \parbox[t]{0.7in}{Red: 22,000\par Blue: 28,000\strut}\\
	\hline	
	
\parbox[t]{1.6in}{J1244+1129\par RA: 12:46:40.37; Dec: 11:13:02.93 \strut} &  3.153 & 2.635 &  19.50$\pm$0.12 & $5400$ & 3500-9,400 & \parbox[t]{0.7in}{Red: 22,000\par Blue: 28,000\strut}\\
		\hline
\parbox[t]{1.6in}{J1614+1448\par RA: 16:14:58.34; Dec: 14:48:36.97 \strut} &  2.548 & 2.236 &  19.75$\pm$0.10 & $5400$ & 3500-9,400 & \parbox[t]{0.7in}{Red: 22,000\par Blue: 28,000\strut}\\		
		\hline
	\end{tabular}
\end{table*}

\begin{figure*}
\centering
  \begin{tabular}{@{}cc@{}}
   \includegraphics[width=.43\textwidth]{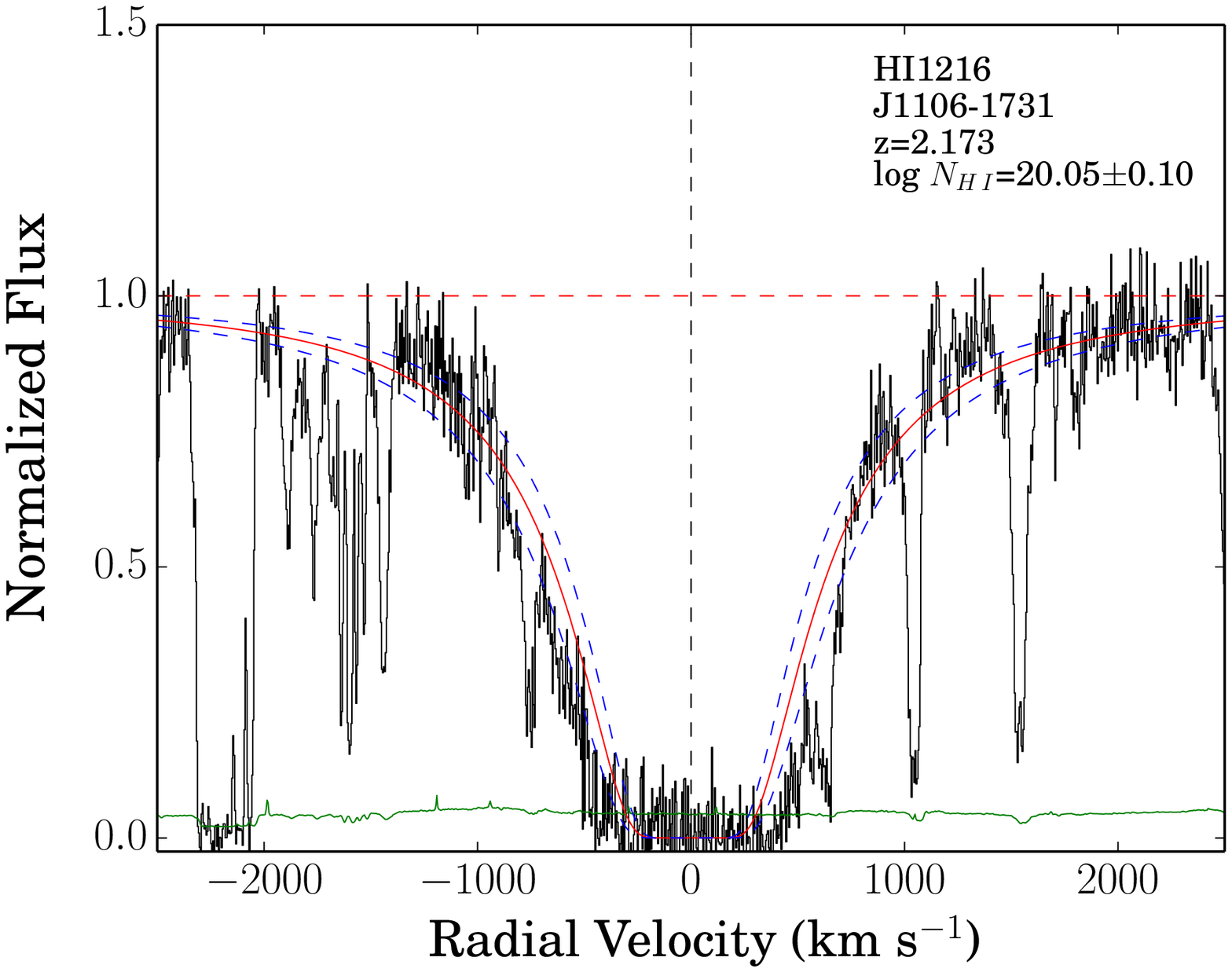} &
   \includegraphics[width=.43\textwidth]{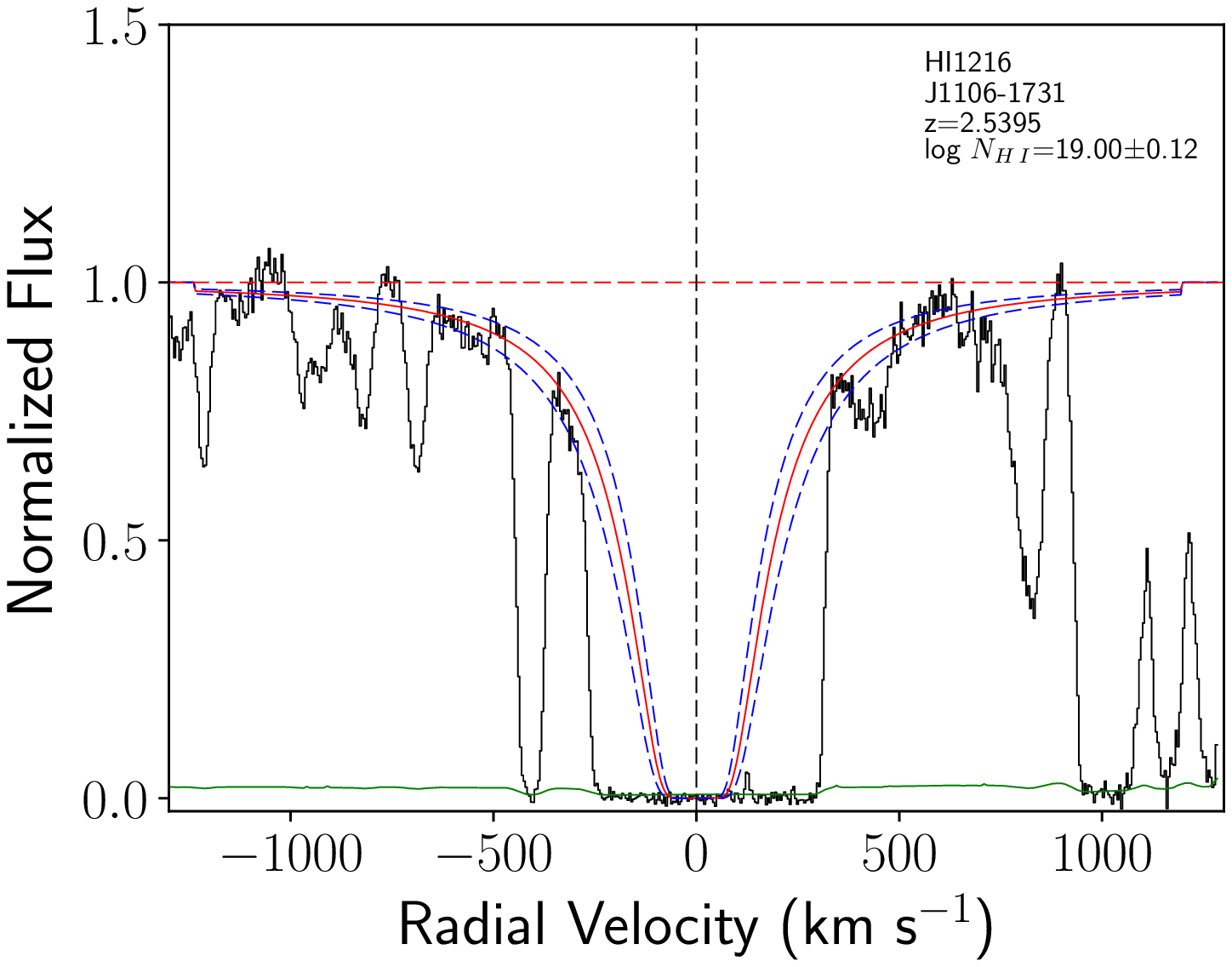}\\
    \includegraphics[width=.43\textwidth]{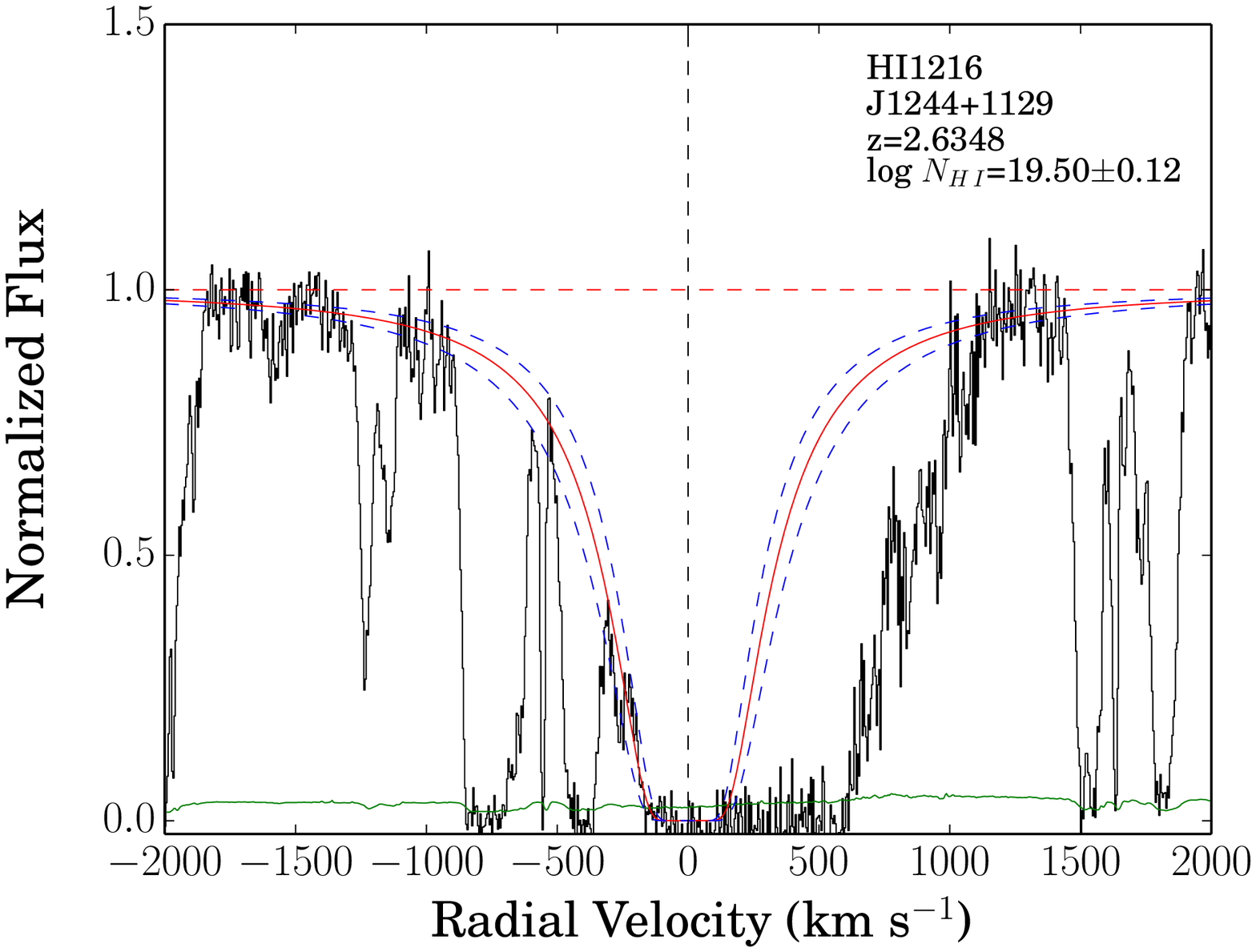} & 
     \includegraphics[width=.43\textwidth]{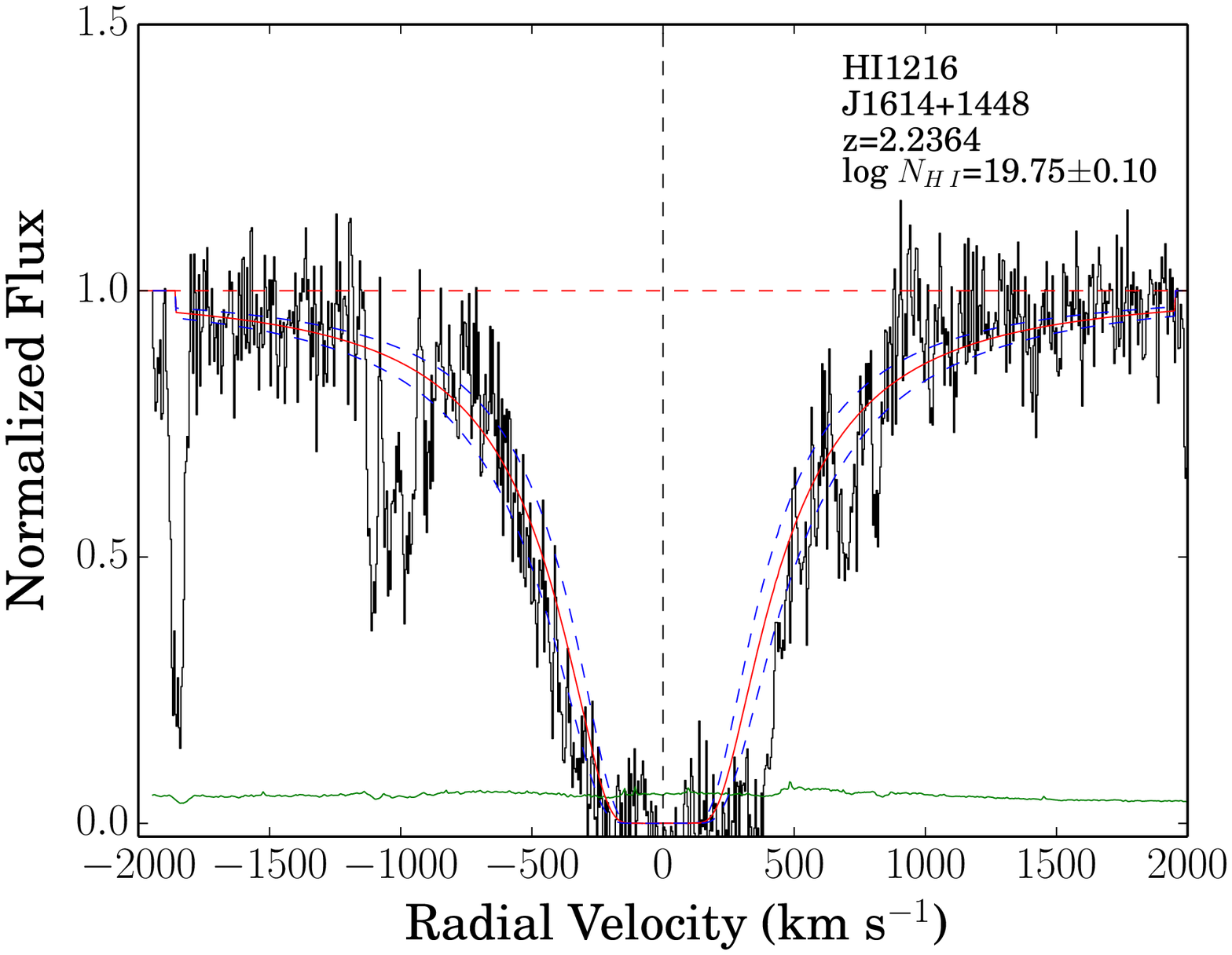} \\ 
  \end{tabular}
  \caption{Plots showing the fitted Voigt profiles for hydrogen Lyman-$\alpha$ for each of our systems. In each case, observed data are shown in black, the solid curve in red shows the best fitted profile with uncertainty in the fits shown by two dashed curves in blue. The continuum level spectrum is represented by the  horizontal dashed line in red. The vertical dashed line in black corresponds to the redshift determined from the fits to the metal lines and the green curve at the bottom of each panel shows the $1\sigma$ error in the continuum-normalized flux. The names of the quasars, redshifts and log $N_{\rm H\,I}$ values of the absorbers are listed in the top right corner of each panel.}
  \label{fig:lyman}
\end{figure*}

\section{Results for individual sub-DLAs}
\label{sec:result}
In this section, we report the results for individual absorbers derived from Voigt profile fitting. We report the column densities of all detected atoms and ions, inferred absolute and relative abundances, and the gas kinematics determined from velocity dispersion measurements. For each system, the H I Lyman-$\alpha$ lines are shown in Fig.~ \ref{fig:lyman}.

\subsection{sub-DLA at $z=2.173$ towards J1106-1731}

\begin{figure}
\hspace*{-0.27in}
\includegraphics[scale=0.48]{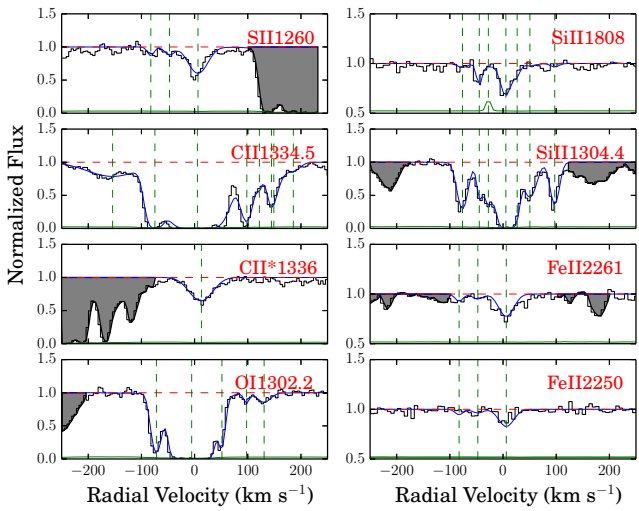}
\caption{Plots showing Voigt profile fits for metal lines for the sub-DLA at $z=2.173$ towards J1106-1731. In each panel, observed data are shown in black, the solid curve in blue shows the fitted profile and the continuum level is represented by the horizontal dashed line in red. The vertical dashed lines in green are centered at the different velocity components and the green line at the bottom of each panel shows the $1\sigma$ error in the continuum-normalized flux. The regions shaded in grey are caused by either telluric absorption or absorption from other systems not associated with the sub-DLA.}
    \label{fig:metals_2173}
\end{figure}

The sight line to J1106-1731 probes a sub-DLA at a redshift of z$=$2.173.  For this system, we determine the H I column density to be log N$_{\rm H\, I}$=$20.05\pm0.10$. We analyzed the metal lines in this system by performing Voigt profile fitting for S II $\lambda$1260, Si II $\lambda$1808, Si II $\lambda$1304, Fe II $\lambda$2250, and Fe II $\lambda$2261. While the O I $\lambda$1302 and C II $\lambda$1334 lines were heavily saturated, we were able to estimate lower limits on O I and C II column densities. The C II$^{\star}$ $\lambda$1336 line was also detected which together with the lower limit on C II column density, allowed us to estimate the cooling rate as well as a limit on the electron density. Fig. \ref{fig:metals_2173} shows the Voigt profile fits for the metal lines and the results from the fits are shown in Table \ref{tab:voigt_metals_2.173}. This system is a metal-rich sub-DLA with [S/H]$=-0.50\pm0.11$. \\

\begin{table*}
	\centering
	\caption{Results from Voigt profile fitting in the z$_{\rm abs}=2.173$ sub-DLA towards J1106-1731.}
	\label{tab:voigt_metals_2.173}
	\begin{tabular}{ccccccccccc} 
	\hline
		 \hline	
		   z  & b$_{\rm eff}$ (km s$^{-1})$  & log $N_{\rm O\, I}$  & log $N_{\rm Si\, II}$& log $N_{\rm Fe\, II}$ & log $N_{\rm S\, II}$ & log $N_{\rm C\, II}$ & log $N_{\rm C\, II\star}$ \\
		  \hline
                 \hline
                 $2.172627$ & $5.54\pm1.14$ &       ...                   & $13.98\pm0.03$&   $13.98\pm0.17$ &$13.6\pm0.20$\\
                 $2.172999$ & $10.58\pm2.09$ &  ...                      &$14.72\pm0.19$& $13.93\pm0.21$&$13.7\pm0.18$\\
                 
                  $2.173208$ & $9.17\pm1.09$ &  ...                      &$13.84\pm0.03$&                            &                         \\
                 
	        $2.173565$  & $24.13\pm1.63$ &  ...                      & $14.98\pm0.05$& $14.78\pm0.04$& $14.58\pm0.04$\\

	        $2.173781$  & $3.31\pm0.63$ &  ...                      & $14.23\pm0.16$&                              &                          \\
	        
	        $2.174034$  & $14.44\pm2.65$ &  ...                      & $13.86\pm0.03$&                              &                          \\
	        	
	        $2.174408$  & $6.26\pm2.14$ &       ...                  & $13.85\pm0.04$&      ...                  &...\\
	       	       	\hline
		Total log N & ... & $>15.34$& $15.29\pm0.06$ & $14.89\pm0.04$ &$14.67\pm0.04$ & $>15.11$ &$13.67\pm0.20$\\
		\hline
		                   [O/H]& [Si/H]  &  [Fe/H]  & [S/H] & [Si/S] & [C/H] \\
		\hline
		                   $>-1.40$& $-0.27\pm0.12$ & $-0.66\pm0.11$ &$-0.50\pm0.11$& $0.23\pm0.06$ & $>-1.37$\\
		\hline
		
		\end{tabular}
\end{table*}

\subsection{sub-DLA at $z=2.539$ towards J1106-1731}

\begin{figure}
\hspace*{-0.27in}
\includegraphics[scale=0.49]{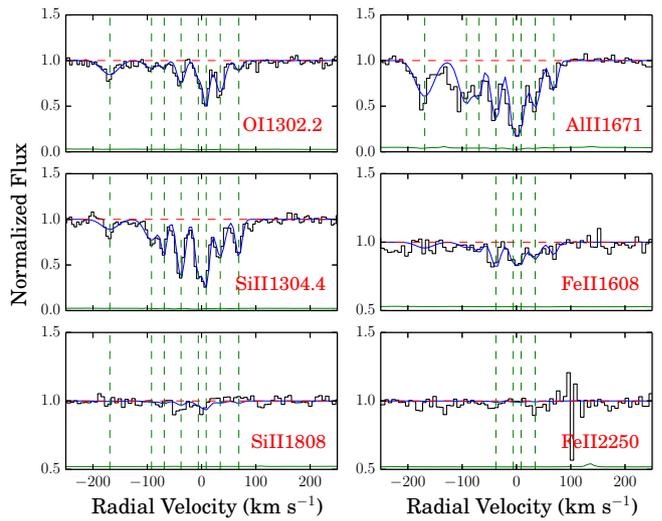}
\caption{Same as Fig. \ref{fig:metals_2173} but showing the Voigt profile fits for the metal lines in the sub-DLA at $z=2.539$ towards J1106-1731.}
    \label{fig:metals_2539}
\end{figure}

\begin{figure}
\hspace*{-0.27in}
\includegraphics[scale=0.48]{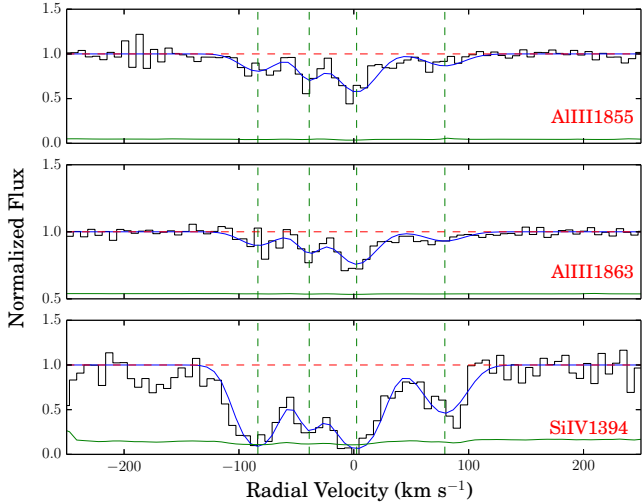}
\caption{Same as Fig. \ref{fig:metals_2539} but showing the Voigt profile fits for the higher ions in the sub-DLA at $z=2.539$ towards J1106-1731.}
    \label{fig:high_2539}
\end{figure}

The sight line to J1106-1731 probes a second sub-DLA at a redshift of z$=$2.539. For this absorber, we determine the H I column density to be log $N_{\rm H I} = 19.00 \pm 0.12$. We analyzed this system by performing Voigt profile fitting for O I $\lambda$1302, Si II $\lambda$1304, Si II $\lambda$1808, Fe II $\lambda$1608, Fe II $\lambda$2250, Al II $\lambda$1671, Al III $\lambda$1855, Al III $\lambda$1863, and Si IV $\lambda$1394. Fig. \ref{fig:metals_2539} and Fig. \ref{fig:high_2539} show the Voigt profile fits for the lower and higher ions respectively. The results from the fits are shown in Table \ref{tab:voigt_metals_2.539} and Table \ref{tab:voigt_metals_2.539_high} respectively. This system has an ionization-corrected metallicity of [O/H]$=-1.27\pm0.12$ (See section~\ref{sec:photo}  below for details about  ionization corrections).\\ 

\begin{table*}
	\centering
	\caption{Results from Voigt profile fitting in the z$_{abs}=2.539$ sub-DLA towards J1106-1731.}
	\label{tab:voigt_metals_2.539}
	\begin{tabular}{ccccccccc} 
	\hline
		 \hline	
		   z  & b$_{\rm eff}$ (km s$^{-1})$  & log $N_{\rm O\, I}$  & log $N_{\rm Si\, II}$& log $N_{\rm Fe\, II}$ & log $N_{\rm Al\, II}$  \\
		  \hline
                 \hline
                 $2.537499$& $20.06\pm2.09$ &  $13.58\pm0.08$& $13.16\pm0.11$  &                               &  $12.40\pm0.07$\\
                 $2.538431$& $15.10\pm2.14$   &  $13.18\pm0.18$& $13.41\pm0.06$   &                              &  $12.41\pm0.06$\\
                 $2.538660$& $2.27\pm0.14$     &  $12.82\pm0.30$& $13.48\pm0.11$   &                             &  $12.19\pm0.27$ \\
                 $2.539030$& $6.98\pm2.14$    &  $13.50\pm0.08$& $13.80\pm0.03$  &  $13.24\pm0.12$  &  $12.53\pm0.08$\\
                 $2.539411$& $4.16\pm1.14$     &  $13.27\pm0.10$& $13.75\pm0.05$   &  $13.05\pm0.19$ &  $12.85\pm0.19$\\
                 $2.539578$& $3.92\pm1.14$   &  $13.87\pm0.07$& $14.05\pm0.08$  &  $13.08\pm0.18$  &  $12.78\pm0.21$\\	
	        $2.539893$& $9.11\pm1.63$    &  $13.72\pm0.06$& $13.55\pm0.04$  &  $13.09\pm0.16$  &  $12.35\pm0.07$\\			       
	        $2.540291$& $4.35\pm1.14$     &  $13.04\pm0.14$& $13.45\pm0.06$  &                              &  $12.06\pm0.17$\\
	       	       	\hline
		Total log N                    & ...        & $14.39\pm0.03$& $14.56\pm0.03$ & $13.72\pm0.08$     &$13.42\pm0.07$ \\
		\hline
		
		\end{tabular}
\end{table*}

\begin{table*}
	\centering
	\caption{Results from Voigt profile fitting for different higher ions in the z$_{abs}=2.539$ sub-DLA towards J1106-1731.}
	\label{tab:voigt_metals_2.539_high}
	\begin{tabular}{ccccccc} 
	\hline
		 \hline	
		   z  & b$_{\rm eff}$ (km s$^{-1})$ &  log $N_{\rm Al\, III}$ & log $N_{\rm Si\, IV}$  \\
		  \hline
                 \hline
                 $2.53849$& $18.77\pm1.14$   &  $12.46\pm0.08$& $13.68\pm0.11$ \\
                 $2.53902$& $10.90\pm2.09$   &  $12.48\pm0.07$& $13.22\pm0.16$  \\	
	        $2.53951$& $20.41\pm1.63$   &  $12.90\pm0.03$& $13.76\pm0.12$ \\		
	        $2.54041$& $21.3\pm2.14$     &  $12.33\pm0.11$& $13.22\pm0.12$ \\
	       	       	\hline
		Total log N & ...                          & $13.20\pm0.03$& $>14.14$ \\
		\hline
		
		\end{tabular}
\end{table*}

\subsection{sub-DLA at $z=2.635$ towards J1244+1129}

\begin{figure}
\includegraphics[scale=0.57]{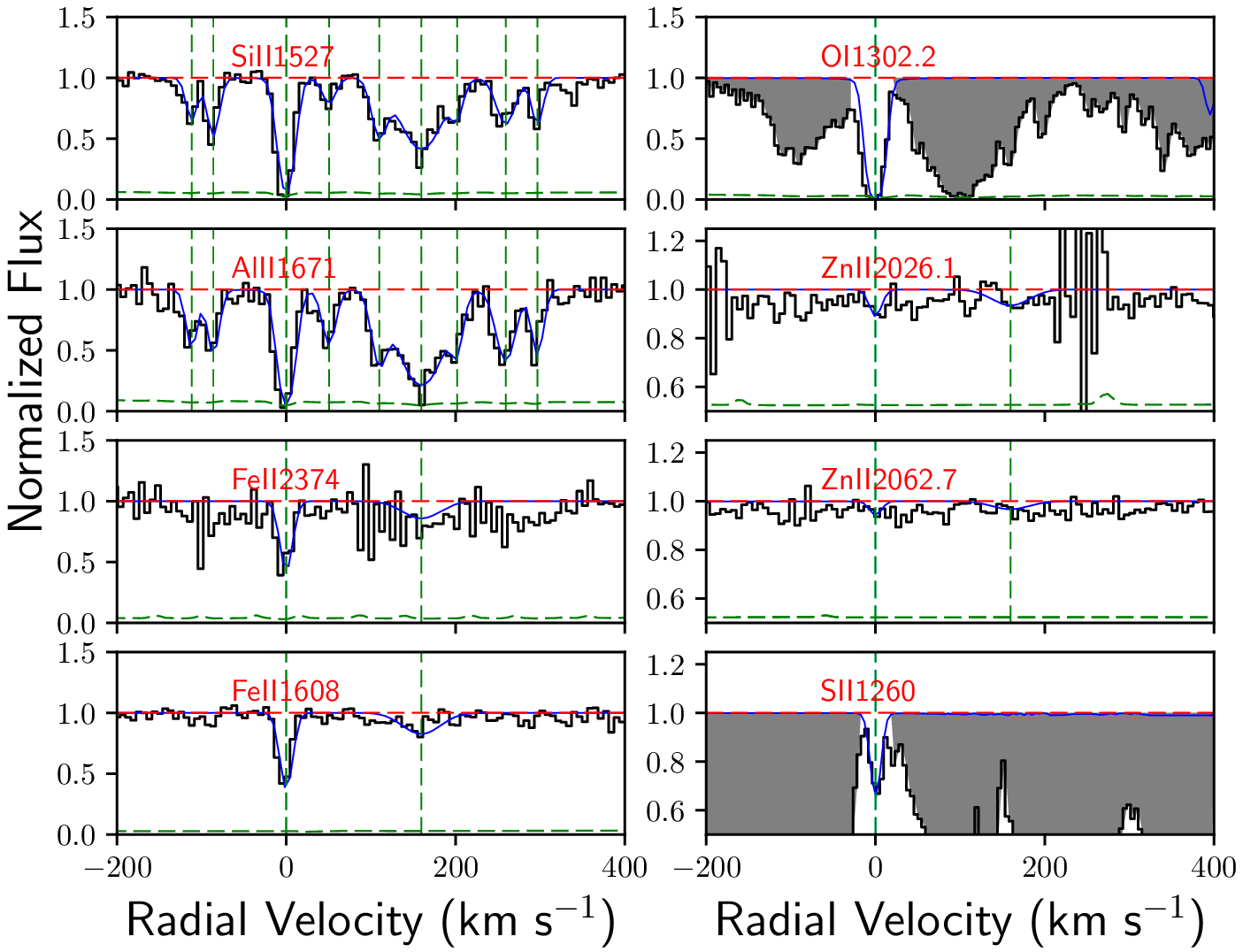}
\caption{Same as Fig. \ref{fig:metals_2173} but showing the Voigt fits for metal lines for the sub-DLA at $z=2.635$ towards J1244+1129.}
    \label{fig:metals_2635}
\end{figure}

\begin{figure}
\includegraphics[scale=0.45]{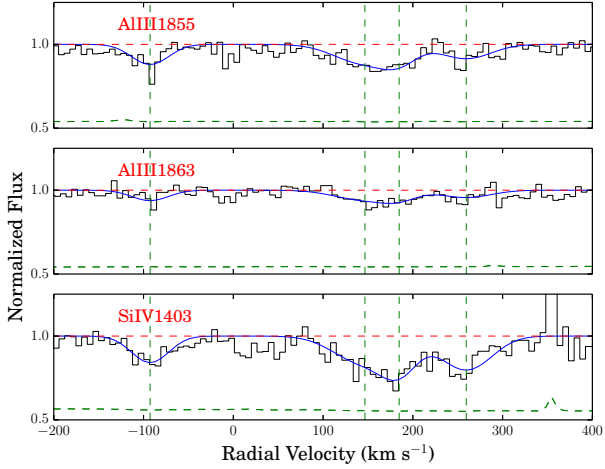}
\caption{Same as Fig. \ref{fig:metals_2635} but showing the Voigt profile fits for the higher ions in the sub-DLA at $z=2.635$ towards J1244+1129.}
    \label{fig:high_2635}
\end{figure}

The sight line to J1244+1129 probes a sub-DLA at a redshift of z$=$2.635. For this absorber, we determine the H I column density to be log $N_{\rm H I} = 19.50 \pm 0.12$. We performed Voigt profile fitting for O I $\lambda$1302, Si II $\lambda$1527, Fe II $\lambda$1608, Fe II $\lambda$2374, Al II $\lambda$1671, S II $\lambda$1260, Zn II $\lambda$2026.1, Zn II $\lambda$2062.7, Al III $\lambda$1855, Al III $\lambda$1863, and Si IV $\lambda$1394. Fig. \ref{fig:metals_2635} and Fig. \ref{fig:high_2635} show the Voigt profile fits for the detected atoms and ions. The results from the fits are shown in Table \ref{tab:voigt_metals_2.635} and Table \ref{tab:voigt_metals_2.635_high}. We note that Zn II $\lambda$2026.1 and Zn II $\lambda$2062.7 absorptions are very weak. However, the ionization-corrected value of [Zn/H]$=0.40\pm0.12$ is consistent with those of [O/H], [Si/H], and [Al/H] (see Table \ref{tab:metals_ic_2.635}). \\

\begin{table*}
	\centering
	\caption{Results of Voigt profile fitting for different elements in the z$_{abs}=2.635$ sub-DLA towards J1244+1129.}
	\label{tab:voigt_metals_2.635}
	\begin{tabular}{cccccccc} 
	\hline
		 \hline	
		   z             &   b$_{\rm eff}$ (km s$^{-1})$   &  log $N_{\rm Si\, II}$  &  log $N_{\rm Al\, II}$   & log $N_{\rm Fe\, II}$   & log $N_{\rm Zn\, II}$   & log $N_{\rm O\, I}$ & log $N_{\rm S\, II}$ \\
		  \hline
                 \hline
                 $2.633463$ & $3.10             $                        &  $13.27\pm0.09$       & $12.43\pm0.15$         &                                       &                                      &\\
                 $2.633773$ & $4.08             $                        &  $13.47\pm0.07$       & $12.41\pm0.11$         &&&\\
                 $2.634817$ & $5.37             $                        &  $15.18\pm0.11$       & $14.15\pm0.16$         &$14.00\pm0.06$            &$11.91\pm0.15$             &$>16.59$ & $>14.08$\\            
                 $2.635429$ & $6.71             $                        &  $12.95\pm0.09$       & $12.22\pm0.07$         &                                       &                                      &\\
                 $2.636147$ & $8.36             $                        &  $13.42\pm0.05$       & $12.42\pm0.06$         &&&\\                          
                 $2.636751$ & $29.3             $                        &  $13.95\pm0.02$       & $13.03\pm0.02$         &$13.62\pm0.06$            &$12.14\pm0.14$             &\\         
                 $2.637263$ & $3.66             $                        &  $13.10\pm0.10$       & $12.33\pm0.13$         &                                      &             &\\
                 $2.637958$ & $12.25           $                        &  $13.38\pm0.04$       & $12.50\pm0.04$         &&&\\              
                 $2.638413$ & $3.83             $                        &  $13.29\pm0.08$       & $12.46\pm0.11$         & &&\\                                                                                         
                 \hline
                 Total LogN  &                                                 &  $>15.24$       & $>14.23$         &$14.15\pm0.05$            &$12.34\pm0.02$             &$>16.59$ &$>14.08$ \\	   
		\hline
		                   [O/H]& [Si/H]  &  [Fe/H]  & [Al/H] & [Zn/H] & [S/H]\\
		\hline
		                   $>0.40$& $>0.23$ & $-0.85\pm0.13$ &$>0.28$&$0.28\pm0.12$ & $>-0.54$\\
		\hline
		
		\end{tabular}
\end{table*}

\begin{table*}
	\centering
	\caption{Results of Voigt profile fitting for different higher ions in the z$_{abs}=2.635$ sub-DLA towards J1244+1129.}
	\label{tab:voigt_metals_2.635_high}
	\begin{tabular}{ccccccc} 
	\hline
		 \hline	
		   z  & b$_{\rm eff}$ (km s$^{-1})$ & log $N_{\rm Si\, IV}$ & log $N_{\rm Al\, III}$ \\
		  \hline
                 \hline
                 $2.633693$& $27.15\pm1.14$    &  $12.96\pm0.08$& $12.37\pm0.06$ \\
	        $2.636593$& $44.75\pm1.63$    &  $13.19\pm0.08$& $12.56\pm0.07$ \\		
	        $2.637055$& $25.82\pm2.14$    &  $13.02\pm0.10$& $12.21\pm0.12$ \\
	        $2.637961$& $37.34\pm2.09$    &  $13.21\pm0.06$& $12.34\pm0.08$  \\	
	       	       	\hline
		Total log N & ...                          & $13.71\pm0.04$& $12.99\pm0.04$ \\
		\hline
		
		\end{tabular}
\end{table*}

\subsection{sub-DLA at $z=2.236$ towards J1614+1448}

\begin{figure}
\includegraphics[scale=0.45]{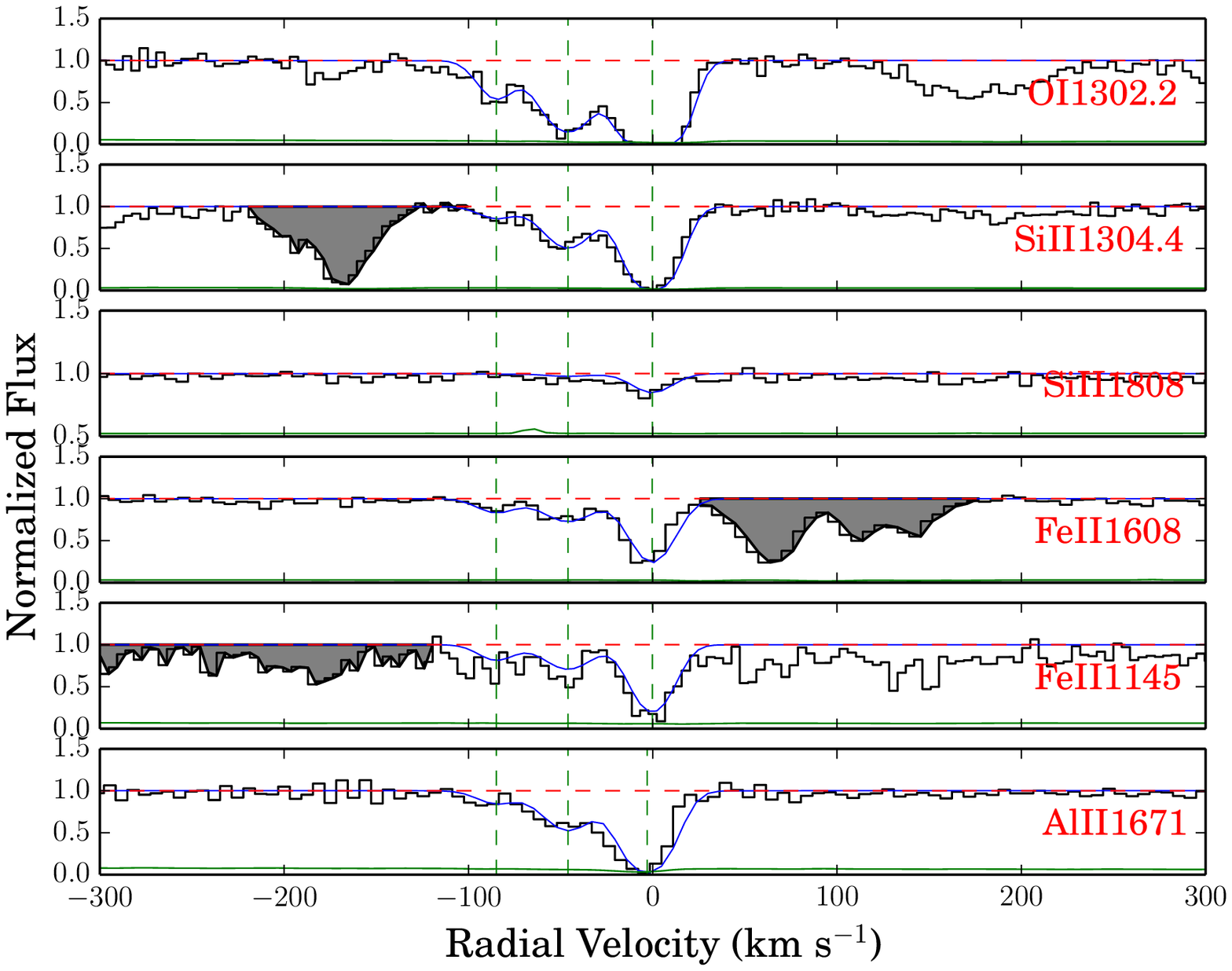}
\caption{Same as Fig. \ref{fig:metals_2173} but showing the Voigt fits for the absorber at $z=2.236$ towards J1614+1448.}
    \label{fig:metals_2236}
\end{figure}

\begin{figure}
\hspace*{-0.27in}
\includegraphics[scale=0.48]{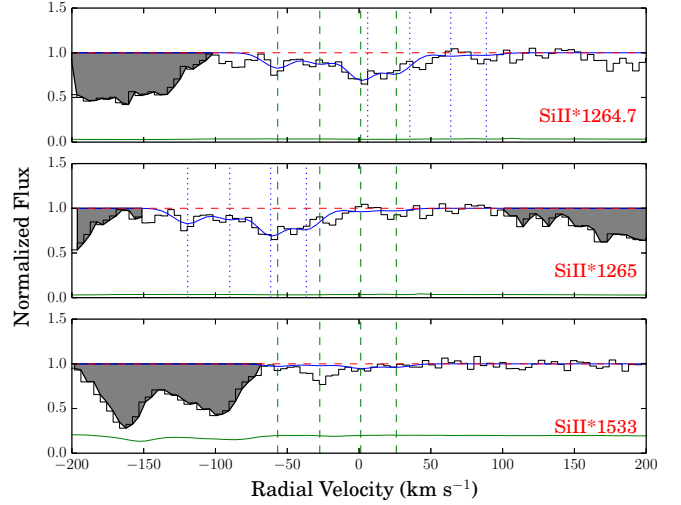}
\caption{ Velocity plots for Si II$^{\star}$ $\lambda$1264.7, Si II$^{\star}$ $\lambda$1265, and Si II$^{\star}$ $\lambda$1533 lines for the absorber at $z=2.236$ in the sight line to J1614+1448. The vertical dashed lines in green show the different velocity components included in the profile fits. The dotted lines in blue in the upper panel show the components corresponding to Si II$^{\star}$ $\lambda$1265 and the dotted lines in blue in the middle panel show the components corresponding to Si II$^{\star}$ $\lambda$1264.7.  The absorption in the shaded regions are not associated with our system.}
    \label{fig:sistar_2236}
\end{figure}

\begin{figure}
\includegraphics[scale=0.45]{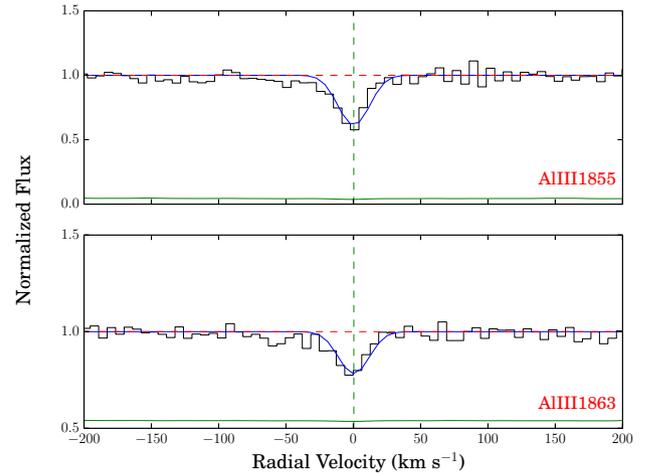}
\caption{Same as Fig. \ref{fig:metals_2236} but showing the velocity plots for higher ions in the sub-DLA at $z=2.236$ towards J1614+1448.}
    \label{fig:high_2236}
\end{figure}

The sight line to J1614+1448 probes a sub-DLA at a redshift of z$=$2.236. For this absorber, we determine the H I column density to be log $N_{\rm H I} = 19.75 \pm 0.10$. We performed Voigt profile fitting for O~I $\lambda$1302, Si II $\lambda \lambda$ 1304, 1808, Fe II $\lambda \lambda$ 1144, 1608, Al II $\lambda$ 1671, Al III $\lambda$ 1855, Al III $\lambda$ 1863, Si II$^{\star}$ $\lambda$1264, and Si II$^{\star}$ $\lambda$1265. As Si II$^{\star}$ $\lambda$1533 appears to be heavily blended with unrelated absorption, we fitted Si II$^{\star}$ $\lambda$1264 and Si II$^{\star}$ $\lambda$1265 together to estimate the Si II$^{\star}$ column density. We used the column densities and Doppler parameters derived from fitting the Si II$^{\star}$ $\lambda$1264 and Si II$^{\star}$ $\lambda$1265 to compare the implied Si II$^{\star}$ $\lambda$1533 profile with the data. The Si II$^{\star}$ column density together with the Si II column density, allowed us to put constraints on the electron densities. Fig. \ref{fig:metals_2236}, Fig. \ref{fig:sistar_2236}, and Fig. \ref{fig:high_2236} show the Voigt profile fits for the detected atoms and ions. The results from the fits are listed in Table \ref{tab:voigt_metals_2.236} and in Table \ref{tab:voigt_high_2.236}. This system is a fairly metal-rich sub-DLA with ionization-corrected metallicity of [O/H]$>-0.84$. \\


\begin{table*}
	\centering
	\caption{Results of Voigt profile fits for elements in the z$_{abs}=2.236$ sub-DLA towards J1614+1448.}
	\label{tab:voigt_metals_2.236}
	\begin{tabular}{cccccccccc} 
	\hline
		 \hline	
		   z                                   & b$_{\rm eff}$ (km s$^{-1})$& log $N_{\rm Al\, II}$  & log $N_{\rm O\, I}$  & log $N_{\rm Si\, II}$& log $N_{\rm Fe\, II}$ & log $N_{\rm Si\, II\star}$ \\
		  \hline
                 \hline
                 $2.23547 $                     & $11.43\pm2.09$                 & $11.74\pm0.27$     &$13.94\pm0.15$    & $13.08\pm0.29$   &$13.28\pm0.26$  &                           \\
                 $2.23589 $                     & $15.79\pm0.09$                 & $12.43\pm0.22$     &$14.55\pm0.09$    & $13.85\pm0.08$   &$13.62\pm0.14$  & $12.06\pm0.28$\\
                 $2.23624$                      & $ 11.49\pm2.28$                &                                &                              &                              &                             &$11.96\pm0.34$\\
                 $2.23639$                      & $11.42\pm1.14$                 &                                &                              &                              &                             & $12.33\pm0.18$ \\
                 $2.23687$                      & $12.29\pm2.28$                 &  $13.20\pm0.25$    &$15.56\pm0.16$    & $14.65\pm0.08$   & $14.21\pm0.08$  &$12.22\pm0.22$\\	        	        
	       	       	\hline
		Total log N                      & ...                                        & $>13.28$                & $>15.61$   & $14.72\pm0.07$   & $14.35\pm0.07$   & $12.77\pm0.12$\\
		\hline
		                   [O/H]       & [Si/H]                &  [Fe/H]               & [Si/O]              & \\
		\hline
		                   $>-0.83$& $-0.54\pm0.12$ & $-0.90\pm0.12$ &$<0.29$&\\
		\hline
		
		\end{tabular}
\end{table*}

\begin{table*}
	\centering
	\caption{Results of Voigt profile fits for higher ions in the z$_{abs}=2.236$ sub-DLA towards J1614+1448.}
	\label{tab:voigt_high_2.236}
	\begin{tabular}{ccc} 
	\hline
		 \hline	
		   z  & b$_{\rm eff}$ (km s$^{-1})$& log $N_{\rm Al\, III}$  \\
		  \hline
                 \hline
                 $2.236344$&$13.21\pm1.1$ & $12.70\pm0.02$ \\
       		\hline
		\end{tabular}
\end{table*}

\section{Discussion}
\label{sec:discussion}
We now examine what implications our results have for various aspects of chemical enrichment processes for the ISM/CGM of galaxies. We compare our results with those from the literature and discuss the trends in metallicity evolution, dust, relative abundances, and constraints on electron densities. Moreover, we discuss the photoionization, star formation rate density, and search for molecules in sub-DLAs.

\subsection{Photoionization}
\label{sec:photo}
Ignoring effects of ionization can lead to errors in the derived element abundances for the systems that consist of both neutral gas (H I regions) as well as ionized gas (H II regions). The low H I column density systems such as Lyman limit systems are heavily affected by the ionization effects. However, for high H I column density systems, ionization effects are either negligible or small (depending on the H I column density) due to the effect of self-shielding of ionizing photons. Therefore, in general, element abundances studies of DLAs ignore ionization corrections. For sub-DLAs, the H I column densities are much higher than LLSs, but smaller than for DLAs. Therefore, ionization corrections for sub-DLAs may not always be small enough to ignore. Past studies suggest that corrections to element abundances due to ionization are generally $\lesssim$0.2 dex for sub-DLAs \citep*[e.g.][]{Dess et al. 2003, Meiring et al. 2009, Cooke et al. 2011, Som et al. 2015}. To assess the ionization corrections for three sub-DLAs in our sample, we ran the the plasma simulation code Cloudy v. 13.03 \citep[e.g.][]{Ferland et al. 2013}. For the sub-DLA at z$=$2.173 that has relatively high H I column density (log N$_{\rm H\, I}$=$20.05\pm0.10$), we have not performed any photoionization corrections due to non-availability of higher ions to constrain the ionization parameter U = $\frac{\rm n_{\gamma}}{\rm n_{H}}$ (the ratio of the number density of ionizing photons to that of neutral hydrogen).\\

Each absorbing system is considered as a slab of uniform density which is heated by the cosmic microwave background (CMB) radiation and the extragalactic ultraviolet (UV) background radiation at the redshift of the absorber. The UV background, which plays an important role in the ionization state of intergalactic medium, was adopted from \citet{ Khaire 2019}. These models use the latest values of quasar emissivity, star formation rate density of galaxies, dust attenuation, and the distribution of the intergalactic H I gas. In addition, the cosmic ray background was also included in our simulation. The observed $\frac{\rm Al\,III}{\rm Al\, II}$ ratio was used to constrain the ionization parameter U for all the systems. In addition, for two cases, we also used the observed $\frac{\rm Si\,IV}{\rm Si\, II}$ ratio.\\

\begin{figure}
\begin{tabular}{l}
\includegraphics[scale=0.62]{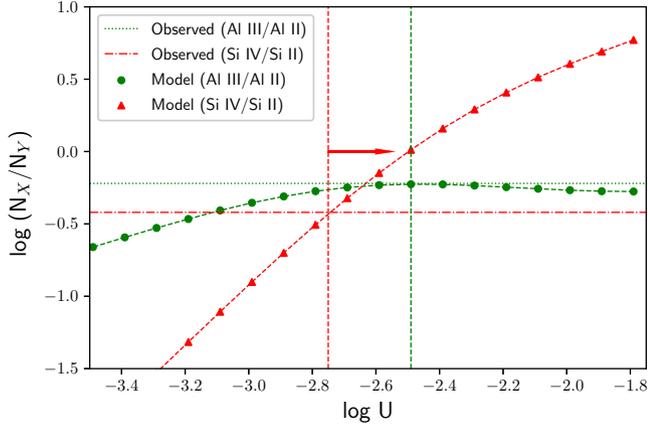}
\end{tabular}
\caption{Results of Cloudy photo-ionization calculations for the absorber at $z=2.539$ in the sight line to J1106-1731, showing the ion ratios as a function of the ionization parameter U. The vertical lines correspond to the value of U implied by the observed $\frac{\rm Si\,IV}{\rm Si\, II}$ and $\frac{\rm Al\,III}{\rm Al\, II}$ ratios (indicated by the horizontal lines in red and blue respectively). The red arrow indicates that log U implied by the observed $\frac{\rm Si\,IV}{\rm Si\, II}$ ratio is a limit consistent with the value determined from the observed $\frac{\rm Al\,III}{\rm Al\, II}$ ratio.}
    \label{fig:photoi}
\end{figure}

For the system at $z=2.539$ with log N$_{\rm H\, I}$=$19.00\pm0.12$, we used grids of models in the range -1.0$>$log n$_{\rm H}$$>$-3, calculating the predicted column density ratio $\frac{\rm Al\,III}{\rm Al\, II}$. Comparing this prediction to the observed Al III/Al II ratio gave log U$=$-2.49 and log n$_{\rm H}$$=$-2.4 (see Fig. \ref{fig:photoi}). The resultant ionization corrections to be applied is given as 
\begin{equation}
\rm correction =  $(log $\frac{\rm N\, X_{total}}{\rm N\, H_{\rm total}}$ - log $\frac{\rm N\, X_{\rm dominant\, ion}}{\rm H\, I}$$)
\end{equation}
where, X is an arbitary element, H$_{\rm total}$=H I + H II, X$_{\rm total}$= X I + X II + X III + ...  ) for various elements were then calculated. This implies ionization corrections of 0.03 dex for O, -0.87 dex for Si, -1.01 dex for Al, and -0.38 dex for Fe (see Table \ref{tab:metals_ic}). The determination of the ionization correction for O was negligible (0.03 dex). This is understandable, because the ionization potential of O I is nearly equal to that of H I. For the other elements, the corrections seem quite high, which is not surprising as this system is in the borderline of being a Lyman limit system and a sub-DLA. Moreover, the ionization parameter determined from the observed $\frac{\rm Si\,IV}{\rm Si\, II}$ ratio was log U$>$-2.75, which is consistent with the value determined from the observed $\frac{\rm Al\,III}{\rm Al\, II}$ ratio.\\

\begin{figure}
\begin{tabular}{l}
\hspace*{-0.27in}
\includegraphics[scale=0.62]{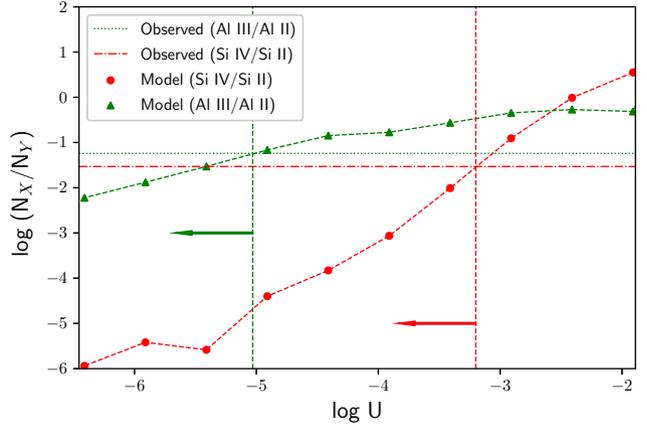}
\end{tabular}
\caption{Same as Fig. \ref{fig:photoi} but showing the results of Cloudy photo-ionization calculations for the absorber at $z=2.635$ in the sight line to J1244+1129. The red and green arrows indicate that log U implied by both the observed $\frac{\rm Si\,IV}{\rm Si\, II}$ and $\frac{\rm Al\,III}{\rm Al\, II}$ ratios are limits, suggesting that the ionization potentials would be even less than the values suggested by the vertical lines and the correction would be even smaller.}
    \label{fig:photoi_2.635}
\end{figure}

\begin{figure}
\begin{tabular}{l}
\hspace*{-0.27in}
\includegraphics[scale=0.62]{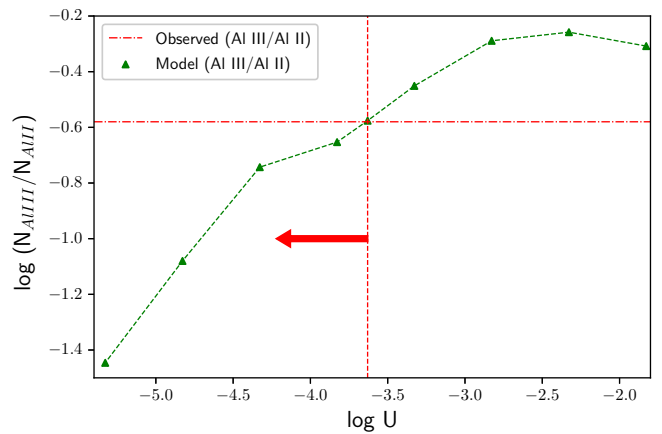}
\end{tabular}
\caption{Same as Fig. \ref{fig:photoi} but showing the results of Cloudy photo-ionization calculations for the absorber at $z=2.236$ in the sight line to J1614+1448. The red arrow indicates that log U implied by both the observed $\frac{\rm Al\,III}{\rm Al\, II}$ ratio is a limit, suggesting that the ionization parameter would be less than the value suggested by the vertical line and the correction would be even smaller.}
    \label{fig:photoi_2.236}
\end{figure}

For the system at $z=2.635$ with log N$_{\rm H\, I}$=$19.50\pm0.12$, we used grids of models in the range 1.5$>$log n$_{\rm H}$$>$-3 to calculate the predicted column density ratios log $\frac{\rm Al\,III}{\rm Al\, II}$ as well as log $\frac{\rm Si\,IV}{\rm Si\, II}$. Comparing the prediction to the observed Al III/Al II ratio gave log U$=$-5.01 and log n$_{\rm H}$$=$0.1 (see Fig. \ref{fig:photoi_2.635}). The ionization corrections are quite small ($\sim \pm 0.02$ dex) for O, Si, Fe, and Al, and modest ($0.12$ dex) for Zn (see Table \ref{tab:metals_ic_2.635}). Since Si II appears to be saturated, the ionization parameter constrained from log $\frac{\rm Si\,IV}{\rm Si\, II}$ is log U $<-$3.1 which is consistent with the estimate from the observed $\frac{\rm Al\,III}{\rm Al\, II}$ ratio. Similarly, for the system at $z=2.236$ with log N$_{\rm H\, I}$=$19.75\pm0.10$, we used grids of models in the range 0.5$>$log n$_{\rm H}$$>$-3, calculating the predicted column density ratio log $\frac{\rm Al\,III}{\rm Al\, II}$. This gave log U$=$-3.63 and log n$_{\rm H}$$=$-1.2 (see Fig. \ref{fig:photoi_2.236}). The ionization corrections for this system are quite small ($\sim \pm 0.1$ dex for Si, and Fe) and the correction for O is again negligible ($\sim0.01$ dex)(see Table \ref{tab:metals_ic_2.236}). Since Al II appears saturated, the ionization parameters for both of these absorbers are likely to be lower than the values listed above and the corrections for absorbers are likely to be even smaller. The ionization corrections for all of these absorbers are listed in Tables \ref{tab:metals_ic}, \ref{tab:metals_ic_2.635}, and \ref{tab:metals_ic_2.236}. It is clear  that the ionization corrections for the elements we use as metallicity indicators are either negligible (for O) or small ($\sim 0.1$ dex for Zn). These results are in agreement with past studies of ionization in sub-DLAs \citep*[e.g.][]{Dess et al. 2003, Meiring et al. 2009, Cooke et al. 2011, Som et al. 2015}.

\begin{table*}
	\centering
	\caption{Total and relative element abundances in the z$_{abs}=2.539$ absorber along the sight line to J1106-1731, before and after ionization correction.}
	\label{tab:metals_ic}
	\begin{tabular}{cccccc} 
	\hline
		 \hline	
		   Element  &  [X/H]$_{\rm NoIC}$  & Correction  & [X/H]$_{\rm IC}$& [X/O]$_{\rm NoIC}$   &  [X/O]$_{\rm IC}$   \\
		  \hline
                O    & $-1.30\pm0.12$    & +0.03  & -1.27$\pm$0.12 & &\\
                Si    & $0.05\pm0.12$     & -0.87   & -0.82$\pm$0.12 & 1.35$\pm$0.04 & 0.45$\pm$0.04\\
                Fe   & $-0.78\pm0.14$    & -0.38   & -1.16$\pm$0.14 & 0.52$\pm$0.08 & 0.11$\pm$0.08\\
                Al    &  $-0.03\pm0.14$   & -1.01   & -1.04$\pm$0.14 & &\\
		\hline
		
		\end{tabular}
\end{table*}

\begin{table*}
	\centering
	\caption{Total and relative element abundances in the z$_{abs}=2.635$ absorber along the sight line to J1244+1129, before and after ionization correction.}
	\label{tab:metals_ic_2.635}
	\begin{tabular}{cccccc} 
	\hline
		 \hline	
		   Element  &  [X/H]$_{\rm NoIC}$  & Correction  & [X/H]$_{\rm IC}$& [X/O]$_{\rm NoIC}$   &  [X/O]$_{\rm IC}$   \\
		  \hline
                O    & $>0.40$                & -0.004  & $>0.40$                & &\\
                Si    & $>0.23$                & -0.02   & $>0.21$                 &  & \\
                Fe   & $-0.85\pm0.13$    & -0.02   & -0.87$\pm$0.13     &  & \\
                Al    &  $>0.28$               & 0.02    & $>0.30$                 & &\\
                Zn   &  $0.28\pm0.12$    & 0.12   & 0.40$\pm$0.12       & &\\
		\hline
		
		\end{tabular}
\end{table*}

\begin{table*}
	\centering
	\caption{Total and relative element abundances in the z$_{abs}=2.236$ absorber along the sight line to J1614+1448, before and after ionization correction.}
	\label{tab:metals_ic_2.236}
	\begin{tabular}{cccccc} 
	\hline
		 \hline	
		   Element  &  [X/H]$_{\rm NoIC}$  & Correction  & [X/H]$_{\rm IC}$& [X/O]$_{\rm NoIC}$   &  [X/O]$_{\rm IC}$   \\
		  \hline
                O    & $>-0.83$   & -0.01  & $>-0.84$ & &\\
                Si    & $-0.54\pm0.11$   & -0.15  & $-0.69\pm0.11$ & $<0.29$    & $<0.15$ \\
                Fe   & $-0.90\pm0.12$   & -0.12  & $-1.02\pm0.12$ & $<-0.07$   &$<-0.17$ \\
                Al    &  $>-0.92$             & 0.09   & $>-0.83$           & &\\
		\hline
		
		\end{tabular}
\end{table*}

\subsection{Metallicity, relative abundances, and dust}
In general, the sub-DLA global mean metallicity seems to be higher than that of DLAs in the redshift range 0$\lesssim$z$\lesssim$3 for which both DLA and sub-DLA observations are available \citep*[e.g.][]{Som et al. 2013, Som et al. 2015}. As sulfur, oxygen, and zinc are weakly depleted on interstellar dust grains, they are believed to probe nearly dust-free or intrinsic metallicity.  The zinc-based metallicity for the absorber at z$=$2.635 along the sight line to J1244+1129 is supersolar ([Zn/H]=0.40$\pm$0.12). Similarly, sulfur-based metallicity for the absorber at z$=$2.173 along the sight line to J1106-1731 is [S/H]=-0.50$\pm$0.11. The oxygen-based metallicities for the absorbers at z$=$2.539 and at z$=$2.236 along the sight lines to J1106-1731 and J1614+1448 are estimated to be [O/H]=-1.27$\pm$0.12 and [O/H]$>-0.84$, respectively. This shows that all of our 4 sub-DLAs are found to be metal-rich based on the weakly depleted elements compared to typical DLAs at comparable redshifts.\\

Si and Fe are found to be under-abundant relative to Zn, with [Si/Zn] > $-0.19$ and [Fe/Zn] = $-1.27$ for the absorber at z$_{abs}$=2.635 along the sight line to J1244+1129. Similarly, Fe is found to be under-abundant relative to S, with [Fe/S]$=-$0.16 for the absorber at z$_{abs}$=2.173 along the sight line to J1106-1731. Such [Si/Zn], [Fe/Zn], and [Fe/S] values can be expected if the refractory elements (elements with higher condensation temperature) such as Si and Fe are depleted onto dust grains more severely than the volatile elements (elements with lower condensation temperature) such as Zn and S. However, it is interesting to note that the Si is overabundant in comparison to S for the absorber at z$=$2.173, with [Si/S] = $0.23 \pm 0.06$. Such a high [Si/S] value is very surprising, as [Si/S] is often either negative or zero depending on whether the dust depletion exists or not.

\subsection{Gas Kinematics}
\citet{Ledoux et al. 2006}, \citet{Moller et al. 2013}, and \citet{Peroux 2003} reported a relation between velocity width vs. metallicity of the absorption-line systems. The velocity width of absorption lines could be related to the gravitational potential well of the absorption system's host galaxy \citep*[e.g.][]{Prochaska et al. 1997b, Haeh 1998, Pontzen et al. 2008} and may be taken as a proxy for the stellar mass. Thus, one potential interpretation of the velocity width vs. metallicity relation is in terms of the stellar mass vs. metallicity relation (MZR) of galaxies, assuming galaxy luminosity scales with the dark matter halo mass. While many previous studies suggest a correlation between mass and metallicity, \citet{Zwaan et al. 2008} show that the velocity width and mass do not correlate well in local analogues of DLAs. We measured the velocity width values for the systems in our sample following the analysis of \citet{Wolfe et al. 1998}. The low-ionization transitions were used to measure the velocity width for all the systems. High-ionization lines are not appropriate as they are more likely to be dominated by large-scale thermal motions in the gas. Moreover, we selected only those transitions that are not so strong as to be affected by saturation, and at the same time, are not so weak as to be undetectable in some velocity components (so as to avoid underestimating the velocity widths).\\

\begin{figure}
\begin{tabular}{l}
\hspace*{-0.27in}
\includegraphics[scale=0.4]{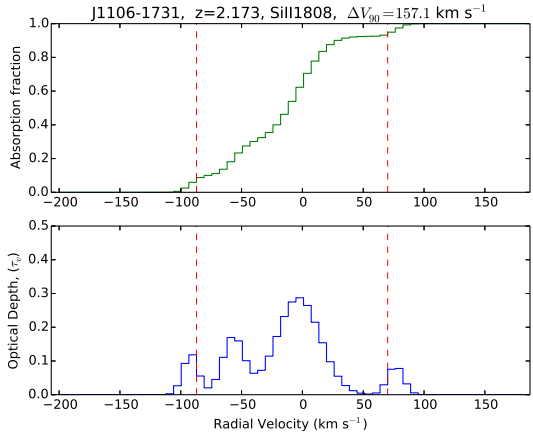}
\end{tabular}
\caption{Upper panel: Absorption fraction vs. radial velocity. Lower panel: Optical depth vs. radial velocity for the sub-DLA at z$_{abs}$ = 2.173 towards J1106-1731. The top panel lists the name of the sight line, redshift of the sub-DLA, name of the transition used, and the corresponding velocity width. The velocity width is estimated as the radial velocity range that contains 90 percent of the cumulative optical depth (i.e., the difference in radial velocity between the two vertical dashed lines in red drawn at 5 percent and 95 percent of cumulative optical depth).}
    \label{fig:vel_2.173}
\end{figure}

\begin{figure}
\begin{tabular}{l}
\hspace*{-0.27in}
\includegraphics[scale=0.4]{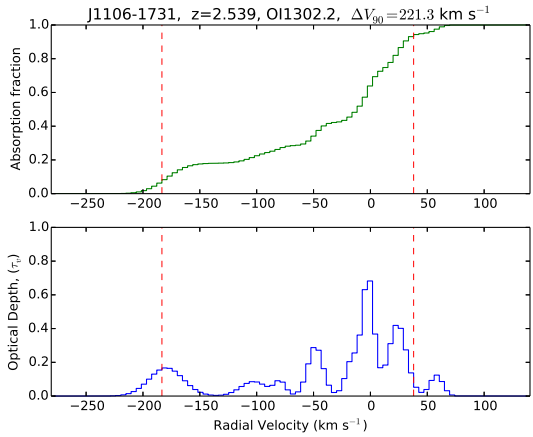}
\end{tabular}
\caption{Same as Fig. \ref{fig:vel_2.173}  but for the sub-DLA at z$_{abs}$ = 2.539 towards J1106-1731.}
    \label{fig:vel_2.539}
\end{figure}

\begin{figure}
\begin{tabular}{l}
\hspace*{-0.27in}
\includegraphics[scale=0.4]{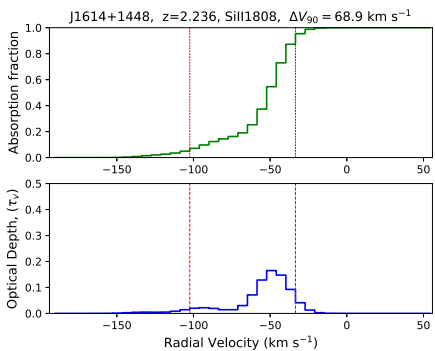}
\end{tabular}
\caption{Same as Fig. \ref{fig:vel_2.173} but  for the sub-DLA at z$_{abs}$ = 2.236 towards J1614+1448.}
    \label{fig:vel_2.236}
\end{figure}

\begin{figure}
\begin{tabular}{l}
\hspace*{-0.27in}
\includegraphics[scale=0.4]{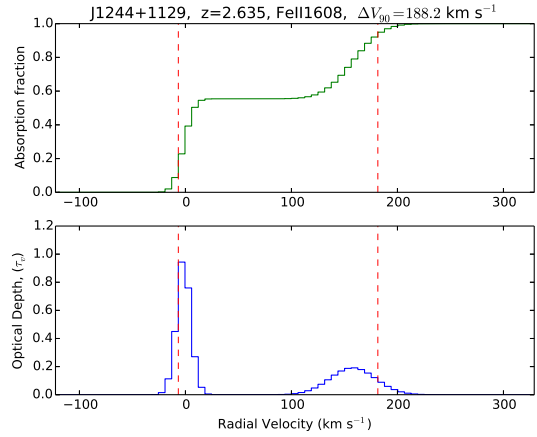}
\end{tabular}
\caption{Same as Fig. \ref{fig:vel_2.173} but  for the sub-DLA at z$_{abs}$ = 2.635 towards J1244+1129.}
    \label{fig:vel_2.635}
\end{figure}

The Si II absorption lines in the z$_{abs}$ = 2.173 and z$_{abs}$ = 2.236 sub-DLAs along the sight lines to J1106-1731 and J1614+1448 are spread over velocity widths of 157 km s$^{-1}$ and 69 km s$^{-1}$, respectively ~(Fig. \ref{fig:vel_2.173} and Fig. \ref{fig:vel_2.236}). The observed values of metallicities for these two systems are $\lesssim2\sigma$ of the expected values based on the metallicity-velocity relation
\begin{equation}
[X/H] = $$(0.88\pm0.08)$ log $\Delta V_{90} - (1.86\pm0.16),$$ 
\end{equation}
observed for lower redshift sub-DLAs \citep[]{Som et al. 2015}. Similarly, the Fe II absorption for the z$_{abs}$ = 2.635 sub-DLA along the sight line to J1244+1129 has a velocity width of 188.2 km s$^{-1}$, which is also consistent with the prediction from the metallicity-velocity relation (see Fig. \ref{fig:vel_2.635}). Moreover, the O I absorption in the z$_{abs}$ = 2.539 sub-DLA along the sight line to J1106+1731 is spread over velocity width of 221 km s$^{-1}$ (see Fig. \ref{fig:vel_2.539}). The metallicity of this absorber ($-1.27 \pm 0.12$) is significantly lower than the value one can expect from the MZR relation for sub-DLAs from \citet{Som et al. 2015}.

\subsection{Constraints on Electron Densities}
The collisional excitation of an electron to the higher energy state is followed by spontaneous radiative de-excitation. By assuming equilibrium between these two processes for a given temperature, fine structure lines such as Si II$^{\star}$ $\lambda \lambda$1264, 1265, 1533 and C II$^{\star}$ $\lambda$ 1335.7 can be used to put constraints on the electron density for an absorber \citep[e.g.][]{Srianand 2000}.
Si II$^{\star}$ absorption is generally detected in gamma-ray burst (GRB) afterglows \citep[e.g.][]{Savaglio 2012}. However, it is not very common in quasar absorbers. Only a few detections of Si II$^{\star}$ exist in DLAs \citep*[e.g.][]{Kulkarni et al. 2012, Noterdaeme et al. 2015}. We detect Si II$^{\star}$ absorption in the sub-DLA at z$_{abs}$ = 2.236 along the sight line to J1614+1448. We performed Voigt profile fitting of Si II$^{\star}$ $\lambda \lambda$1264, 1265, 1533 simultaneously and determined log N$_{\rm Si II^{\star}}$ to be 12.77$\pm$0.12. The Si II collisional excitation rate was assumed to be 
\begin{equation}
$C$_{12}$ = 3.32 $\times$ 10$^{-7}$ (T/10$^{4}$)$^{-0.5}$ $\exp{(-413.4/T)}$ cm$^{3}$ s$^{-1},$$
\end{equation}
and the Si II$^{\star}$ spontaneous radiative de-excitation rate A$_{21}$=2.13$\times10^{-4}$ s$^{-1}$ \citep[e.g.][]{Srianand 2000}. The electron density is then given by 
\begin{equation}
$n$_{e}$ = (N$_{\rm Si II^{\star}}$/ N$_{\rm Si II}$) A$_{21}$/C$_{12}.$$
\end{equation}
As C$_{12}$ depends on temperature, we estimate the electron densities at two different temperatures T = 500 K and T = 7000 K. We obtained n$_{e}$ $=$ 3.63 cm$^{-3}$ and n$_{e}$ $=$ 6.30 cm$^{-3}$, respectively, for T = 500 K and T = 7000 K. These values are much higher than the median value n$_{e}$=0.0044$\pm$0.0028 cm$^{-3}$ found in DLAs \citep[e.g.][]{Neeleman et al. 2015} and also higher than the range of electron density values  (0.007-0.047 cm$^{-3}$) found in the  H$_{2}$-bearing high-$z$ DLAs \citep[e.g.][]{Srianand 2005}. In fact, our above-mentioned estimates of the electron density in the sub-DLA toward J1614+1448 are higher than even the values found in some super-DLAs \citep*[e.g.][]{Kulkarni et al. 2012, Kulkarni et al. 2015, Noterdaeme et al. 2015}, the highest of which is in the range $n_{e} =$ 0.53-0.91 cm$^{-3}$.\\

Moreover, we were able to detect C II$^{\star}$ $\lambda$ 1335.7 for the absorber at z = 2.173 in the sight line to J1106-1731. However, as most of the components of C II $\lambda 1334$ were saturated for this system, we were able to put only an upper limit on the electron density, n$_{e}$.  Again, we estimated the electron density 
\begin{equation}
$n$_{e}$ = (N$_{\rm C II^{\star}}$/ N$_{\rm C II}$) A$_{21}$/C$_{12}$,$
\end{equation}
 by assuming equilibrium between the collisional excitation and the radiative de-excitation of C II, where, A$_{21}$=2.29$\times$10$^{-6}$ s$^{-1}$ \citep[e.g.][]{Nuss 1981}. The collision rate coefficient is given by 
 \begin{equation}
 $C$_{12}$(T)=[8.63$\times$10$^{-6}$ $\Omega_{12}$/(g$_{1}$T$^{0.5}$)]$\exp{(-E_{12}/kT)}$,$
 \end{equation}
  \citep[e.g.][]{Wood 1997}, where, g$_{1}$=2, E$_{12}$=1.31$\times$10$^{-14}$ erg, and the collision strength $\Omega_{12}$ depends on temperature. We were able to put upper limits of n$_{e}$ $<$0.18 cm$^{-3}$ and n$_{e}$ $<$0.58 cm$^{-3}$, respectively, for T = 500 K and T = 7000 K. These values are consistent with the median value $n_{e} = 0.0044 \pm 0.0028$ cm$^{-3}$ found in DLAs \citep[e.g.][]{Neeleman et al. 2015}. 

\subsection{Cooling Rate}
\begin{figure}
\begin{tabular}{l}
\hspace*{-0.27in}
\includegraphics[scale=0.4]{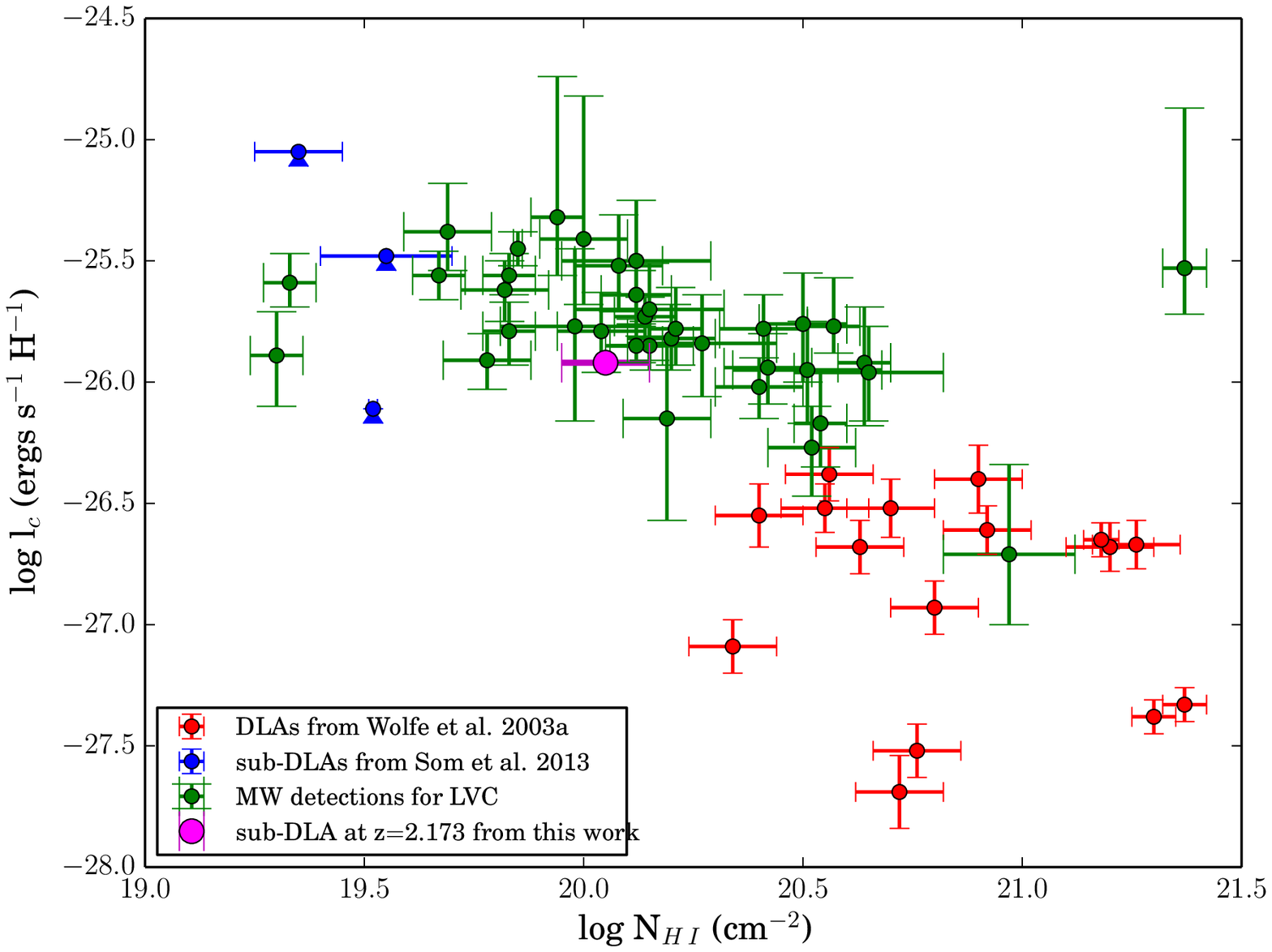}
\end{tabular}
\caption{Cooling rate per H atom, l$_{c}$ inferred from C II$^{\star}$ column density, versus H I column density. The red points denote the sample of QSO DLAs in \citet{Wolfe et al. 2003}. The green points represent the measurements for low-velocity interstellar H I clouds in the Milky Way compiled in \citet{Lehner et al. 2004}. The point in magenta represents a sub-DLA at z=2.173 along the sight line to J1106-1731 from this work. The blue points denote the lower limits for the sub-DLAs from \citet{Som et al. 2013}.}
    \label{fig:sfr}
\end{figure}

\citet{Wolfe et al. 2003} developed a technique to infer the star formation rate per unit area for individual damped Lyman-alpha systems. In this technique, the [C II] 158 $\mu$m cooling rate is inferred from the C II $\lambda$1335.7 absorption line in the neutral gas producing the damped Ly-alpha absorption. The C II$^{\star}$ $\lambda$ 1335.7 transition arises from the excited $^{2}$P$_{3/2}$ state in C$^{+}$. A spontaneous photon decay of the $^{2}$P$_{3/2}$ state to the $^{2}$P$_{1/2}$ state results in [C II] 158 $\mu$m emission, which is the principal coolant of neutral gas in the Galactic ISM \citep[e.g.][]{Wright 1991}. At thermal equilibrium, the cooling rate equals the heating rate, which makes it possible to calculate the star formation rate per unit area. The cooling rate can be expressed as 
\begin{equation}
$$l_{c}=\frac{N_{C II\star}E_{ul}A_{ul}}{N_{H I}},$$
\end{equation} 
where E$_{ul}$ and A$_{ul}$ denote the energy and coefficient for spontaneous photon decay for the transition from the $^{2}$P$_{3/2}$ state to $^{2}$P$_{1/2}$ \citep[e.g.][]{Pottasch et al. 1979}. Using A$_{ul}$=2.29$\times10^{-6}$, we estimate l$_{c}$=1.20$\times10^{-26}$ erg s$^{-1}$ per H atom for the absorber at z $=$ 2.173 in the sight line to J1106-1731, suggesting a higher star formation rate density in this sub-DLA than the typical star formation rate density for DLAs at similar redshifts from \citet{Wolfe et al. 2003}. The cooling rate versus H I column density data for the sub-DLA from this work along with the corresponding measurements for DLAs from \citet{Wolfe et al. 2003}, sub-DLAs from \citet{Som et al. 2013}, and for interstellar clouds in the Milky Way adopted from \citet{Lehner et al. 2004} are plotted in Fig. \ref{fig:sfr}. While our value of the cooling rate is lower than that for some other sub-DLAs and for low-velocity H I gas in the Milky Way ISM, it is higher than the cooling rate for all the DLAs from \citet{Wolfe et al. 2003}. 

\subsection{Search for CO molecules}

\begin{figure*}
\centering
  \begin{tabular}{@{}cc@{}}
   \includegraphics[width=.43\textwidth]{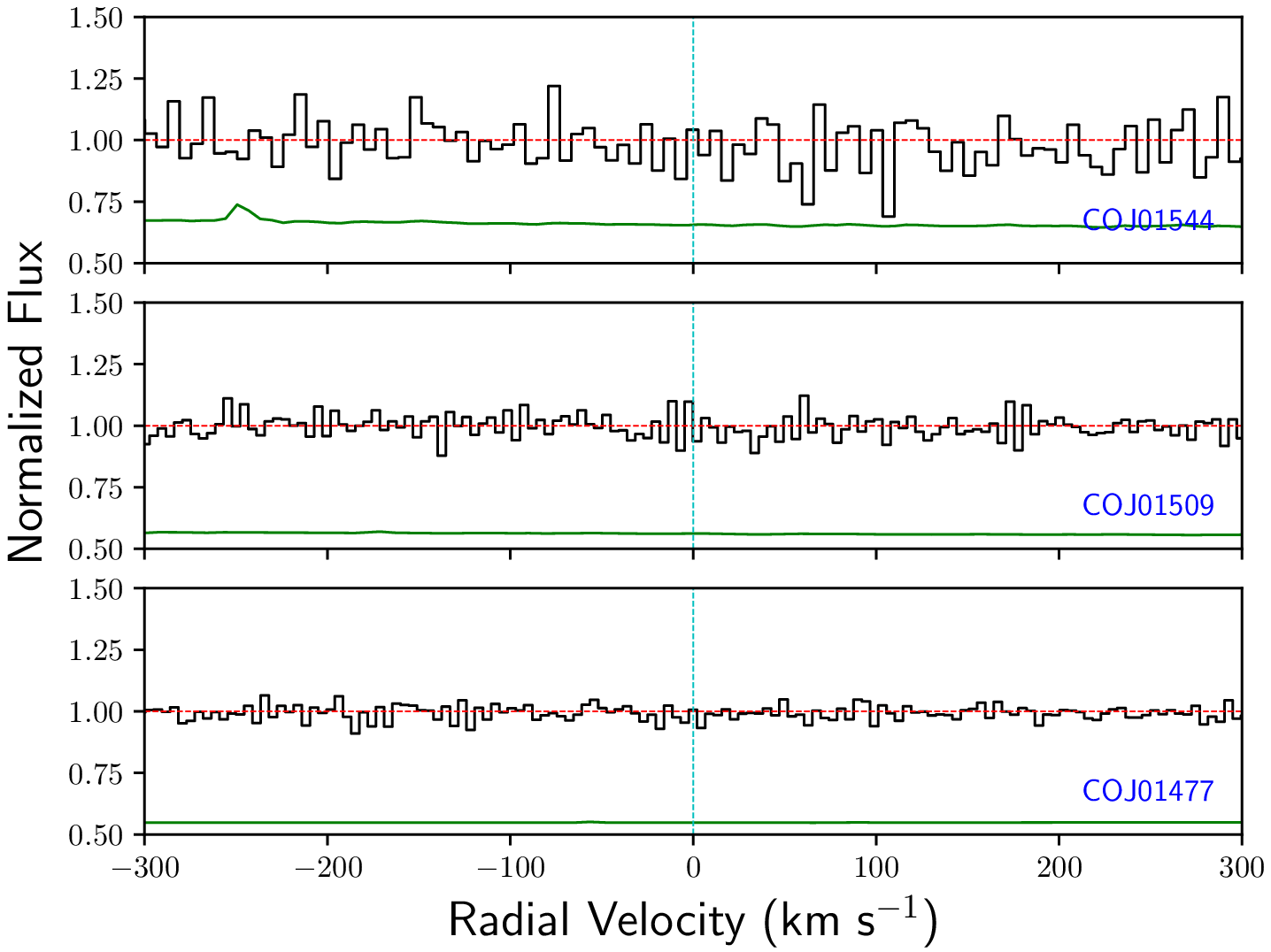} &
   \includegraphics[width=.43\textwidth]{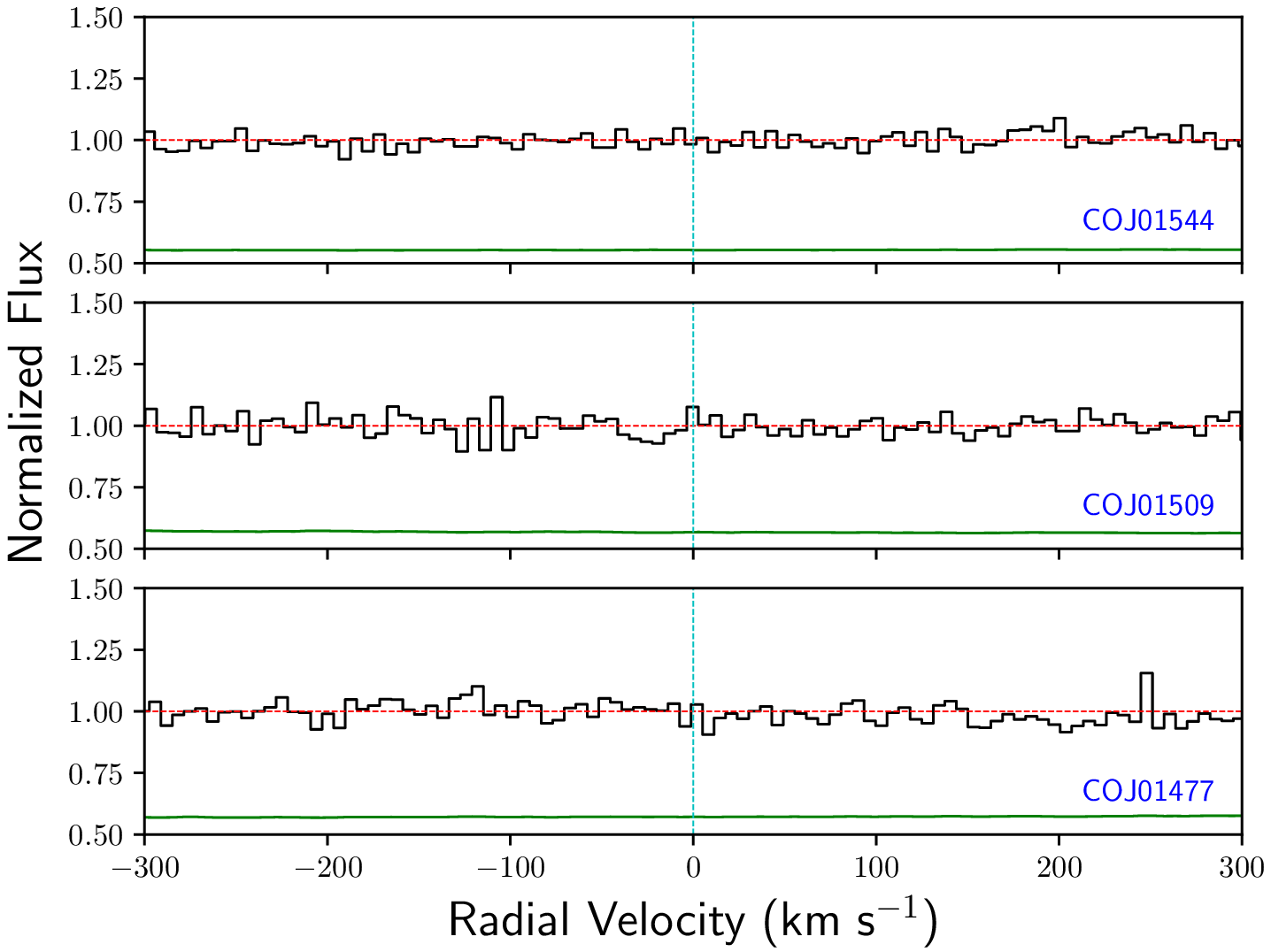}\\
    \includegraphics[width=.43\textwidth]{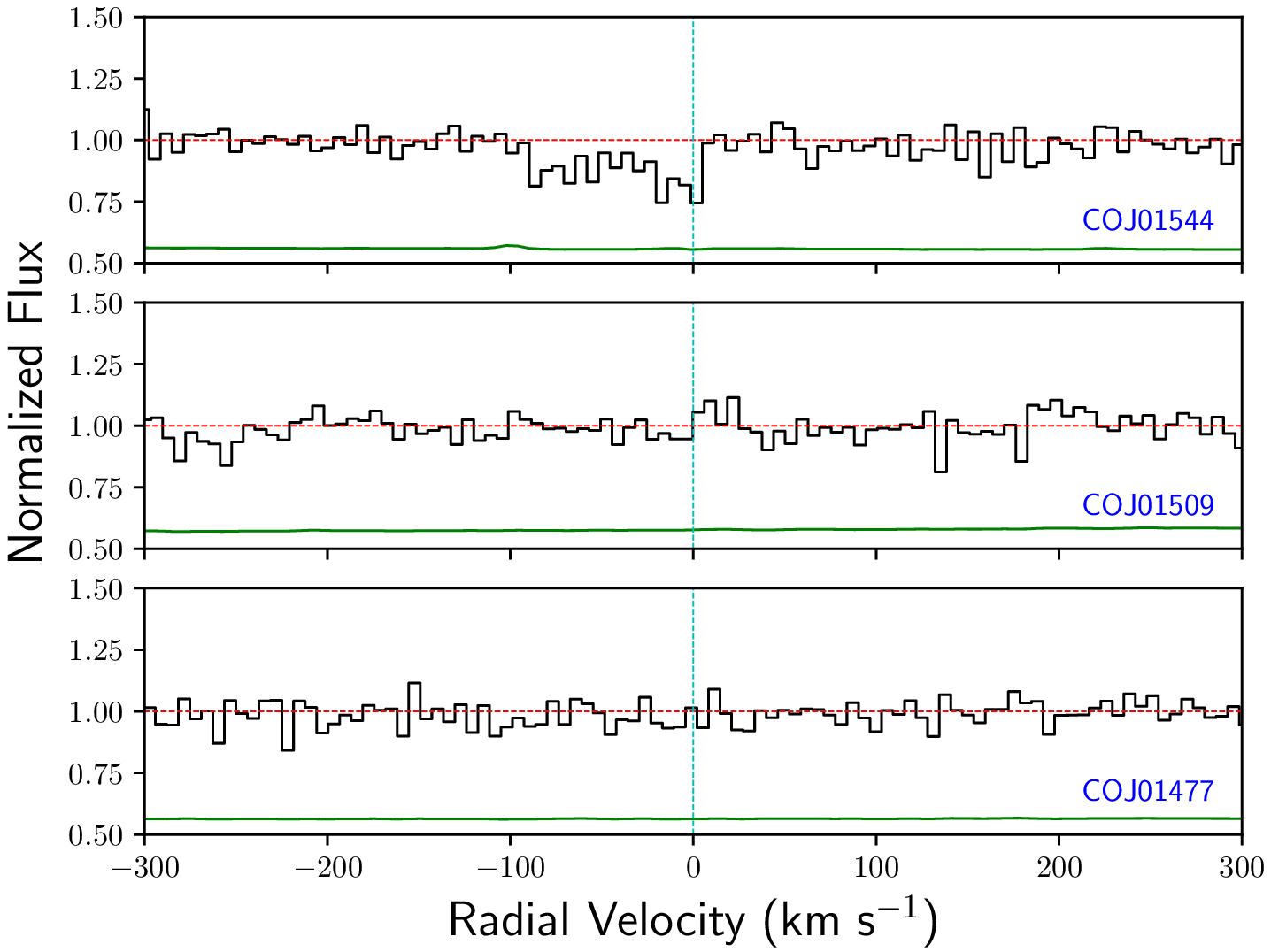} & 
     \includegraphics[width=.43\textwidth]{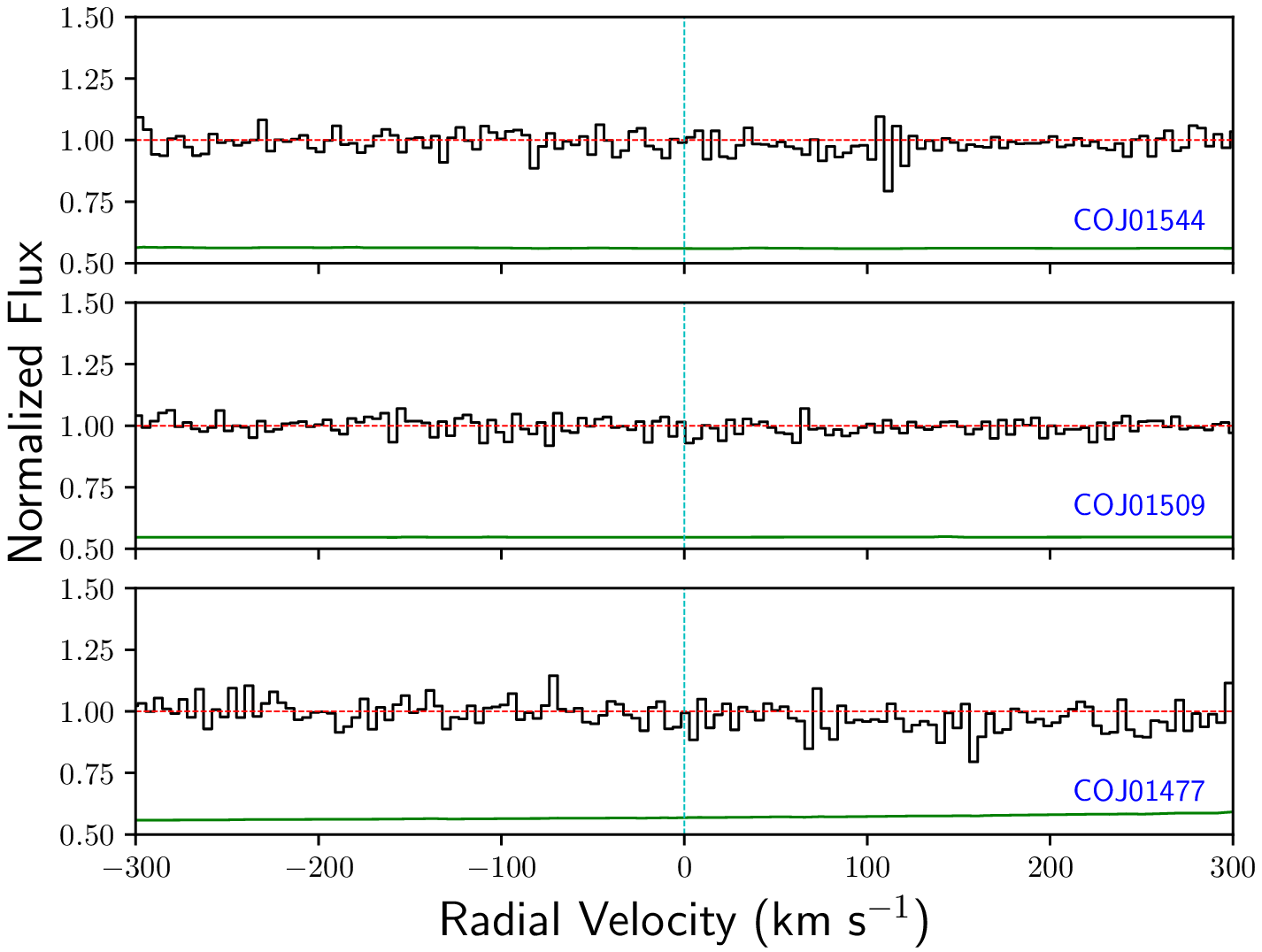} \\ 
  \end{tabular}
  \caption{Plot showing the positions of molecular lines of CO for all of the absorbers. The top two plots represent systems at z$=$2.173 (left) and z$=$2.539 (right), and the bottom two plots represent systems at z$=$2.635 (left) and z$=$2.236 (right), respectively.}
  \label{fig:mole}
\end{figure*}

\begin{figure}
\begin{tabular}{l}
\hspace*{-0.27in}
\includegraphics[scale=0.5]{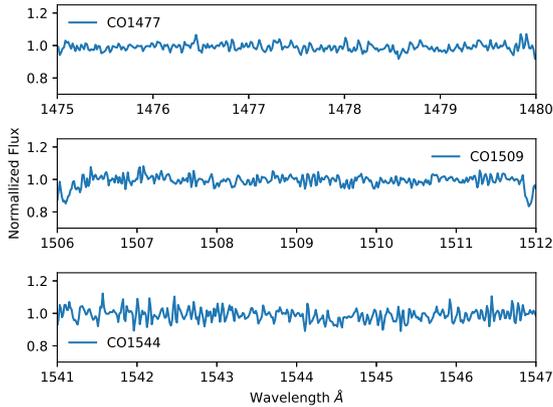}
\end{tabular}
\caption{Plot showing the positions of molecular lines of CO after stacking the rest frame spectra of all the absorbers together.}
    \label{fig:mol_rest}
\end{figure}

Most of the Lyman and Werner band absorption lines of H$_{2}$ for all the absorbers in our sample are not covered by the spectra. While a few of these lines are covered, the S/N in this region is very low and, furthermore, these lines are blended with the Lyman-alpha forest. For all of our absorbers, several lines of CO were covered, but they were not detected. We estimate 3$\sigma$ upper limits of log N(CO, J=0) $<$  13.87, log N(CO, J=0) $<$  13.17, and log N(CO, J=0) $<$  13.08, respectively, from the non-detections of absorption from the $J=0$ level in the CO AX 0-0, 1-0,  and 2-0 bands near 1544, 1510, 1478 {\AA}, respectively. These limits were calculated by estimating the average SNR per pixel adjacent to the expected positions of the molecular lines. Fig. \ref{fig:mole} shows the velocity structures at the positions of the above-stated CO bands and Table \ref{tab:mole} lists the corresponding 3$\sigma$ upper limits on the CO column density for $J=0$ from the non-detections of absorption from the $J=0$ level in these bands. There is a slight hint of absorption in the CO A-X 0-0 band near 1544.45 {\AA} for the $z=2.635$ system, but no corresponding absorption (expected to be stronger) in the CO A-X 1-0 and 2-0 bands near 1509.75 and 1477.56 {\AA}, respectively. Moreover, we made a rest-frame spectral stack of the four absorbers in order to increase the SNR and look for weak absorption. However, we do not notice any significant absorption, as seen in Fig. \ref{fig:mol_rest}, which shows the velocity plots for the CO A-X 0-0, 1-0,  and 2-0 bands near 1544, 1510, 1478 {\AA}, respectively, after making the rest-frame stacking.\\

\begin{figure}
\begin{tabular}{l}
\hspace*{-0.27in}
\includegraphics[scale=0.5]{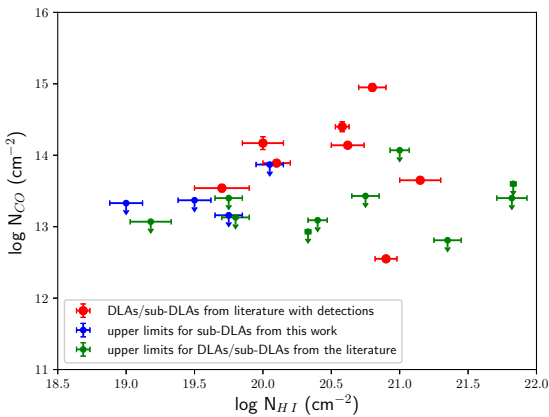}
\end{tabular}
\caption{Plot showing the comparison of log N$_{\rm HI}$ vs. log N$_{\rm COJ0}$ for the sub-DLAs from this work (upper limits) with sub-DLAs/DLAs from literature (including both the limits as well as the detection. The data from literature are compiled from \citet{Srianand 2008}, \citet{Noterdaeme et al. 2011, Noterdaeme et al. 2015, Noterdaeme 2017, Noterdaeme 2018}.}
    \label{fig:HICO}
\end{figure}

CO absorption features in the background quasar provides an extremely useful tool to understand the chemical properties of the galaxies and to put constraints on fundamental physics (e.g., measuring the temperature of the cosmic microwave background radiation). However, detection of CO absorption remains extremely challenging as they require high resolution and high SNR spectra to detect the weak and closely spaced rotational bands of CO. There exists only a handful of CO detections in the literature along the lines of sight to background quasars \citep[e.g.][]{Srianand 2008, Noterdaeme et al. 2011, Noterdaeme et al. 2015, Noterdaeme 2017, Noterdaeme 2018}. The upper limits of CO column densities for our absorbers are lower than most of the detections in the literature (see Fig. \ref{fig:HICO}). This may be understandable as all of our systems have relatively low H I column densities and molecules are generally detected in high column density absorbers. The first detection of such kind was in a DLA at z$_{abs}=$2.4 with log N$_{\rm CO}=13.89$ and log N$_{\rm HI}=20.10$ \citep[e.g.][]{Srianand 2008}. Similarly, \citet{Noterdaeme 2017} reported a DLA at z$_{abs}=$2.4 with log N$_{\rm CO}=14.95$ and log N$_{\rm HI}=20.80$. Both of these systems have either near-solar or super-solar metallicities and depletion patterns similar to those in cold gas in the diffuse ISM. We also note that one of our systems at z$=$2.635 that shows slight hint of absorption in the CO A-X 0-0  band near 1544.45 {\AA}, has similar metallicity and depletion level as the two systems in the literature. However, we can not confirm if this is truly an absorption from CO molecules as no hint of absorption can be seen in the stronger CO A-X 1-0 and 2-0 bands near 1509.75 and 1477.56 {\AA}. Higher S/N and lower wavelength spectra of our sub-DLAs are essential to perform more sensitive searches for H$_{2}$ and CO.\

\begin{table*}
	\centering
	\caption{Limits on column densities of different CO molecular transitions for all the absorbers in this paper.}
	\label{tab:mole}
	\begin{tabular}{ccccccc} 
	\hline
		 \hline	
		   QSO            & z$_{abs}$  & log $N_{\rm CO\,J0\, 1544}$ & log $N_{\rm CO\,J0\, 1509}$&log $N_{\rm CO\,J0\, 1477}$\\
		  \hline
                 \hline
                 $J1106-1731$& $2.173$     &  $<13.87$          & $<13.10$&$<12.68$ \\
	        $J1106-1731$& $2.539$     &  $<13.33$           & $<13.11$&$<13.08$ \\		
	        $J1244+1129$& $2.635$    &  $<13.37^{\star}$& $<13.17$&$<13.03$ \\
	        $J1614+1448$& $2.236$    &  $<13.16$           & $<12.71$&$<12.79$  \\	
		\hline
		
		\end{tabular}
\end{table*}

\section{Conclusions}
\label{sec:conclusion}

In this paper, we have presented high-resolution absorption spectra of four sub-DLAs at 2.173 $<$ z$_{abs}$ $<$ 2.635, increasing the existing sample for undepleted elements in sub-DLAs and thus improving the constraints on the cosmic metal evolution. We find a spread in the metallicities, which range from -1.27 dex to +0.40 dex. These observations suggest that metal-rich sub-DLAs appear at high redshift as well, supporting the conclusion made by \citet{Som et al. 2013}. We are also able to put constraints on the electron density by assuming equilibrium between collisional excitation and spontaneous radiative de-excitation for an assumed gas temperature using fine structure lines of C II$^{\star}$ and Si II$^{\star}$ for two of the sub-DLAs. These values are much higher than the values found in DLAs (including H$_{2}$-bearing DLAs) and even in super-DLAs. We estimate the cooling rate for a sub-DLA at z $=$ 2.173 in the sight line to J1106-1731 using C II$^{\star}$$\lambda$1335.7 line to be l$_{c}$=1.20$\times10^{-26}$ erg s$^{-1}$ per H atom, suggesting higher SFR density in this sub-DLA than the typical SFR density for DLAs at similar redshifts. We also study the metallicity versus velocity dispersion relation for our absorbers and compare the values with those from the literature. Most of the absorbers follow the trend one can expect from the mass versus metallicity relation for sub-DLAs from literature. Finally, although our spectra either do not cover or have very low S/N in the locations of H$_{2}$ lines, we are able to put limits on the column density of CO from the non-detections of various strong electronic transitions. We estimate 3$\sigma$ upper limits of log N(CO, J=0) $<$  13.87, log N(CO, J=0) $<$  13.17, and log N(CO, J=0) $<$  13.08, respectively, from the non-detections of absorption from the $J=0$ level in the CO AX 0-0, 1-0,  and 2-0 bands near 1544, 1510, 1478~ {\AA}. We emphasize the need of higher S/N and lower wavelength spectra of our sub-DLAs to obtain more definitive determinations of H$_{2}$ and CO contents.

\section*{Acknowledgements:}
We thank an anonymous referee for  thoughtful comments that have helped to improved this manuscript. SP, VPK, and DS thank the helpful staff of Las Campanas Observatory for their assistance during the observing runs. SP and VPK gratefully acknowledge support from NASA grant NNX17AJ26G (PI Kulkarni) and NSF grant AST-0908890 (PI Kulkarni). VPK also gratefully acknowledges support from NASA grant 80NSSC20K0887 and NSF grant AST/ 2009811.

\section*{Data Availability} 
The data underlying this article will be shared on reasonable request to the corresponding author.








\appendix



\bsp	
\label{lastpage}
\end{document}